\documentclass[12pt,a4paper]{article}
\usepackage[includeheadfoot,
marginratio={1:1,2:3},
width=525pt,
height=750pt,]{geometry}

\usepackage{amsmath}
\usepackage{amsfonts}
\usepackage{amssymb}
\usepackage{graphicx}
\usepackage{cancel}
\usepackage{empheq}
\usepackage{color}
\usepackage{hyperref}

\numberwithin{equation}{section}
\usepackage{float}
\restylefloat{table}
\restylefloat{figure}
\usepackage[utf8]{inputenc}
\usepackage{amsmath, amsthm, amssymb, amsfonts}
\usepackage[font=small, labelfont=bf]{caption}
\usepackage{amsmath, amsthm, amssymb, amsfonts}
\usepackage{multirow}
\usepackage{multicol}
\usepackage{graphicx}
\usepackage{float}
\usepackage[numbers,sort&compress]{natbib}
\usepackage{enumerate}
\usepackage{amsmath}
\usepackage{amsfonts}
\usepackage{amssymb}
\usepackage{graphicx}
\usepackage{cancel}
\usepackage{empheq}
\usepackage{color}
\usepackage{colortbl}
\definecolor{green2}{cmyk}{0, 1, 0.5, 0.3}
\definecolor{green3}{cmyk}{1, 0.75, 1.0, 0.0}

\definecolor{lightgreen}{cmyk}{0.2, 0, 0.2, 0.2}
\definecolor{lightgray}{cmyk}{0.1,0.2,0,0.1}
\definecolor{lightgray2}{cmyk}{0.4,0.4,0,0.8}
\definecolor{black}{cmyk}{1.0,1.0,1.0,1.0} 

\usepackage{hyperref}
\usepackage{float}
\restylefloat{table}
\restylefloat{figure}
\usepackage{longtable}
\usepackage{lipsum} 

\usepackage{amsfonts}
\usepackage{amssymb}
\usepackage{comment}
\usepackage{graphicx}
\usepackage{amsmath}
\usepackage{dsfont}
\usepackage{float}
\usepackage{amsthm}
\usepackage{slashed}
\usepackage{setspace}
\usepackage[labelformat=simple]{subcaption}

\usepackage{pgfplots}
\usepackage{tikz}
\usepackage{tikz-cd}
\usepackage[compat=1.1.0]{tikz-feynman}
\usepackage{feynmp-auto}
\usetikzlibrary{shapes.misc,shadows,fit,decorations.pathmorphing,positioning,trees,decorations.markings,decorations.pathreplacing,calc,shapes,patterns,arrows,positioning,arrows.meta}
\usepackage{xcolor}
\usepackage{pgfplots}
\usepackage{pgfplotstable}
\usepackage{xcolor}
\usepackage{makecell}
\usepackage{diagbox}
\usepackage{environ}
\usepackage[utf8]{inputenc}
\usepackage{amsmath, amsthm, amssymb, amsfonts}
\usepackage{amsmath, amsthm, amssymb, amsfonts}
\usepackage{multirow}
\usepackage{graphicx}
\usepackage{float}
\usepackage[numbers,sort&compress]{natbib}

\usepackage{enumerate}
\usepackage{enumitem}
\usepackage[capitalize,sort]{cleveref}
\usepackage{hhline}
\usepackage{mathtools}
\usepackage{longtable}
\usepackage{makecell}
\usepackage{tabularx}
\usepackage{cellspace}
\usepackage{alphalph}
\usepackage{lscape}
\usepackage{slashed}
\usepackage{setspace}
\usepackage{pifont}

\crefname{figure}{Figure}{Figures}
\crefname{table}{Table}{Tables}

\def\be{\begin{equation}}
\def\ee{\end{equation}}
\def\bea{\begin{eqnarray}}
\def\eea{\end{eqnarray}}
\def\bes{\begin{subequations}}
	\def\ees{\end{subequations}}

\newcommand{\bmat}{\left(\begin{array}}
	\newcommand{\emat}{\end{array}\right)}

\def\1{{\bf 1}}
\def\2{{\bf 2}}
\def\3{{\bf 3}}
\def\4{{\bf 4}}
\def\6{{\bf 6}}

\newcommand{\beq}{\begin{equation}}
\newcommand{\eeq}{\end{equation}}

\allowdisplaybreaks[2]
\numberwithin{equation}{section}

\pgfplotsset{compat=1.18}

\def\be{\begin{equation}}
\def\ee{\end{equation}}
\def\bea{\begin{eqnarray}}
\def\eea{\end{eqnarray}}
\def\bes{\begin{subequations}}
	\def\ees{\end{subequations}}

\usepackage{multicol}
\usepackage{float}
\usepackage{caption}
\def\mN{\mathcal{N}}

\allowdisplaybreaks[2]
\numberwithin{equation}{section}

\tikzfeynmanset{warn luatex=false}
\begin{document}
	\vspace{1.0cm}
	\begin{center}
		{\Large
			Symmetries and Interactions of $\mathcal{N}=1$ SUGRA: from Constructive and BCFW to 
            KLT formulations}
		\vspace{0.4cm}
	\end{center}
	\vspace{0.35cm}
	\begin{center}
		Dibya Chakraborty$^{a,b}$, J. Lorenzo D\'iaz-Cruz$^a$, Jonathan Reyes P\'erez$^c$, Pablo Ortega Ruiz$^c$ \footnote{Email:  dibyac@physics.iitm.ac.in, jldiaz@fcfm.buap.mx, \\jonathan.reyesper@alumno.buap.mx, pablo.ortegar@alumno.buap.mx}
	\end{center}
		
\vspace{0.1cm}
\begin{center}
{$^a$ Centro Interdisciplinario de Investigacion y Educacion de la Ciencia, BUAP
\protect\\Ciudad Universitaria, Puebla, Pue. M\'exico}\\
\vspace{0.3cm}
{$^b$ Centre for Strings, Gravitation and Cosmology, Department of Physics, Indian Institute of Technology Madras, Chennai 600036, India}\\

\vspace{0.3cm}
{$^c$ Facultad de Ciencias F\'isico - Matem\'aticas Benem\'erita Universidad Aut\'onoma de Puebla  
\protect \\ Apdo. Postal 1364, C.P. 72000, Puebla, Pue. M\'exico}
\end{center}	
\vspace{1cm}
	
\abstract{In this paper, we study the couplings of  the gravity supermultiplet (graviton and gravitino) of minimal $\mathcal{N}=1$ SUGRA following a constructive approach. Firstly, we use the master formula that follows from considering the scaling behavior of the spinor variables under the little group. Secondly, we derive the 4-point couplings using BFCW. Then, we verify these results for the general 3-point interactions that can be  derived using the KLT-type relations, i.e., they can be written as the square of the coupling of the gluons and gluinos. Finally, we consider SUGRA Compton effect for graviton-gravitino. For completeness, we present in the appendix the $\mathcal{N}=1$ SUGRA lagrangian in the 2-component Weyl formalism, including the proofs of SUSY and gauge invariance.}

\clearpage
	
\tableofcontents


\section{Introduction}

The inclusion of gravity at the quantum level has been an area of active research for almost a century. In the infra-red domain, quantum gravity (QG) lies on the same footing as the other three interactions in nature, with gravity being mediated by the massless spin-2 particle, known as the graviton \cite{Feynman:1963ax, DeWitt1967a,DeWitt1967b,DeWitt:1967uc,tHooft:1974toh}. Standard QFT methods such as quantization, renormalization, and regularization of Yang-Mills theories \cite{Hooft:2016gir}, developed during 60's and early 70's gave perspective to our understanding of weakly-coupled gravity. It was observed that pure QG remains divergence free at one-loop but they start to appear at two-loop in pure gravity. However, addition of matter makes the situation worse where the one-loop level includes divergences, leading to the conclusion that QG is non-renormalizable \cite{tHooft:1974toh}.
 The progress towards effective field theories has been proven to be remarkable where one is able to do loop calculations even for non-renormalizable theories, for example chiral perturbation theory \cite{Weinberg:2016kyd}.  In this paradigm, one can now do reliable calculations for perturbative quantum gravity (pQG)  at the loop-level  \cite{Donoghue:1993eb,Donoghue:1994dn,Bjerrum-Bohr:2002aqa}. 


The modern helicity amplitude method of \cite{Mangano:1990by, Dixon:1996wi, Bern:1996je} has been proven to be another cornerstone in pQG. In this formalism, one uses a unified framework where both kinematical variables and wave functions are expressed in terms of Weyl spinors. For instance, using the Parke-Taylor formula \cite{Parke:1986gb,Berends:1987me}, we are able to obtain more concise expressions for the maximal-helicity violating amplitudes for gluons in Yang-Mills theories as well as for gravitons in pQG \cite{Elvang:2013cua}. Not only that, using Parke-taylor one can show that some amplitudes involving gravitons can be written as the square of the corresponding amplitude for gluons, that is $GR=YM\times YM$ \cite{Bern:1999ji}. This result was first obtained at tree-level in string theory \cite{Kawai:1985xq}, known as the KLT relations. An extension of this relation is known as  the double-copy formula or Bern-Carrasco-Johansson (BCJ) relation \cite{Bern:2008qj, Bern:2010ue} which also holds at loop level. Although most progress has been made for massless  theories, authors of \cite{Arkani-Hamed:2017jhn, Diaz-Cruz:2016abv} have also looked at the massive case .

QFT has evolved over the years, and in particular we have now a better understanding 
of the role of renormalizability in QFT.  After the success of QED by the mid XX century, where it was shown possible to deal with the infinities in QFT through the renormalization program, it was considered that the next goal should be to develop a QFT for the strong and weak interactions. This was achieved thanks to the work of 't Hooft and Veltman, who proved that YM with and without spontaneous symmetry breaking (SSB) could be renormalizable, provided for instance that anomalies are cancelled. Later on, we learned from the work
of Wilson, Weinberg, among others, that a QFT could be organized in terms of relevant, irrelevant and marginal operators.
Although, the long-searched renormalizable theories include only one set of operators. For instance, one could add operators
of dimension higher than four to the usual SM lagrangian, but those operators should include inverse powers of an scale 
$\Lambda$, such that the their effects will be suppressed as compared to the dimension-four operators of the SM. 


Within the EFT paradigm it is possible to treat quantum effects (loops) for a non- renormalizable theory (in the traditional sense),
such as pQG. This can be done in a consistent manner by organizing the short- and 
long-distance effects; but short-distance also means high-energy while long-distance means low-energy. Heavy particles require high-energies to be produced, while massless particles can produce effects at low-energies.
Then, one can prove that the effects of massive particles (high energies) appear as changes in the values of  the coefficients of the operators of the effective lagrangian. On the other hand, the effects of massless particles (low-energies)  can be of a universal nature, i.e. they  are associated with loop-induced non-local terms and are parameter-free.  Here the only massless particle to be considered is the graviton.

Following the usual approach of pQG, one expands the metric ($g_{\mu\nu}$) upon a Minkowski 
background ($\eta_{\mu\nu}$), i.e. $g_{\mu\nu}= \eta_{\mu\nu} + \kappa h_{\mu\nu}$, with the fluctuations ($h_{\mu\nu}$)
identified as the graviton. Then, in order to derive the graviton self-interactions one  substitute this expansion in
the curvature invariant terms making up the EH action.  Notice that the metric is 
dimensionless, $[g_{\mu\nu}]=[\kappa h_{\mu\nu}]=0$, then the mass-dimension of the graviton field is $[h_{\mu\nu}]=1$, 
and $[\kappa]=-1$. Being the derivative of the metric, the Christoffel symbols ($\Gamma$) have mass dimension $[\Gamma]=-1$, while the Riemann tensor, the Ricci tensor and the Ricci scalar $R$ (being the second derivative of the metric), have mass dimensions $[R]=2$, i.e. $\Gamma = O(p)$, while $R= O(p^2)$.  Thus,  the matrix elements of pQG can be organized as an  expansion in energy.

In addition, we also present the $\mathcal{N}=1$  supergravity action in two-components Weyl spinors. We start with the action below and show that it is invariant under global supersymmetry transformations and diffeomorphisms, as shown in the appendix (\ref{appA},\ref{appB})
\begin{align}
S_{\text{SUGRA}}\,&=S_{EH}+S_{RS},\nonumber\\
&= - \frac{1}{2\kappa^2} \int d^4 x \left((   R^L_{\mu \nu} - \frac{1}{2}  \eta_{\mu \nu} R^L         ) h^{\mu \nu}+ \kappa^2 \epsilon^{\mu \nu \rho \sigma} (   \widetilde{\psi}_{\mu} \Bar{\sigma}_{\nu} \partial_{\rho} \psi_{\sigma} - \chi_{\mu} \sigma_{\nu} \partial_{\rho} \widetilde{\chi}_{\sigma}    )\right)
\end{align}

 where $\kappa^2= 8\pi G_N$, and $G_N$ is Newton's gravitational constant. SUGRA is beautiful but complicated \cite{Freedman:1976xh, Wess:1992cp, Freedman:2012zz}, and modern approach in QFT can help to elucidate its structure, both symmetries and interactions \cite{Elvang:2013cua}.


One of the most crucial  ingredients in quantum field theory is the S-matrix encompassing all the scattering information that can occur in the universe with each matrix element denoting a scattering process. Using perturbative methods, it is possible to derive Feynman diagrams from matrix elements \cite{Peskin:1995ev}. Over the years computing amplitudes of interactions by means of Feynman rules seemed irreplaceable but this point of view may change due to on-shell methods which includes helicity spinor formalism to describe massless particles, BCFW recursion relations to name a few \cite{Arkani-Hamed:2012zlh,Britto:2004ap,Britto:2005fq}.

The paper is organised as follows. In section \ref{section2}, we start by applying the master formula for the 3-point interactions from constructive approach to the graviton-gravitino (h - $\widetilde{h}$).  The explicit expressions depend on  the  allowed helicity combinations of the involved particles. Section \ref{section3} includes the study of the 4-point coupling for two gravitons and two gravitinos, derived using BFCW recursion relations. In section \ref{section4}  we turn to traditional Feynman rules and KLT relations to verify the above derivation. Finally, in section \ref{section5} we consider the graviton-gravitino Compton effect to apply the formulation, namely we verify that the amplitudes from each approach coincides. Verifications of gauge and SUSY invariances of $N=1$ SUGRA are left to the appendix.

\section{Constructive approach for 3-point interactions}\label{section2}

The constructible program \cite{Benincasa:2007xk}, also known as the \emph{bootstrap} program  \cite{Cheung:2017pzi} has changed the way we approach a QFT problem.  Lagrangians, Feynman diagrams and Feynman rules can all be replaced in this program --  using the S-matrix, one can derive the fundamental interaction from the general properties of amplitudes and spinors. Within this approach, spinors are considered as the fundamental kinematical variables to write down the amplitudes of massless particles \cite{Elvang:2013cua}. 

To start with, we express the $4-$momenta $p_{\mu}$ as a $2\times 2$ matrix, we write $ p_{\alpha \dot{\alpha}} = \sigma^{\mu}_ {\alpha \dot{\alpha}}  p_{\mu}$ ($\alpha=1,2$ and $\dot{\alpha}=\dot{1},\dot{2}$), 
with the help of the Pauli matrices $\sigma^{\mu}$. We know that for massless particles, we get $\det (p)=0$ and the momentum matrix can be expressed in terms of Weyl spinors as $ p_{\alpha \dot{\alpha}} = \chi_{\alpha}   \tilde{\chi}_{\dot{\alpha}}$. Further, it might be convenient to utilise the Dirac notation where we write $\chi_{\alpha} (p_i)  \to | i \rangle$ and $\tilde{\chi}_{\dot{\alpha}} (p_i)  \to [ i | $, both for the momentum and for the 
wave functions associated with external fermions. We define the inner product among these fermions as $\langle i j \rangle = \chi_i^\alpha \chi_{j\alpha}$ and $ [i j ] = \tilde{\chi}_{i \dot{\alpha}} \tilde{\chi}_j^{\dot{\alpha}}$, so that they can be simplified to:
$\langle j i \rangle [i j ] = 2 p_i \cdot p_j$. Similarly, the polarisation vectors in terms of spinors can be written as: $\epsilon^+_{\alpha \dot{\alpha}} = -\sqrt{2} \eta_{\alpha} \tilde{\chi}_{\dot{\alpha}} / \langle \eta \chi \rangle$, and
$\epsilon^-_{\alpha \dot{\alpha}} = -\sqrt{2} \chi_{\alpha} \tilde{\eta}_{\dot{\alpha}} / [\tilde{\chi} \tilde{\eta}]$, where $\eta, \tilde{\eta}$ denote two reference  spinors that are unphysical and thus they should vanish at the end of a calculation. The end result represents a unified framework where both our kinematics and dynamics are described with the same type of variables. One standard practice in the constructive program for QFT is to attempt to express all the $n-$point amplitudes in terms of the $3-$point amplitudes as in \cite{Cheung:2017pzi}. 

\begin{figure}[H]
    \centering

\tikzset{every picture/.style={line width=0.75pt}} 

\begin{tikzpicture}[x=0.75pt,y=0.75pt,yscale=-1,xscale=1]

\draw    (121,98.5) -- (185,99) ;
\draw    (201,105.75) -- (245,143.5) ;
\draw    (202,91.5) -- (248,50.5) ;
\draw   (185,99) .. controls (185,94.17) and (189.48,90.25) .. (195,90.25) .. controls (200.52,90.25) and (205,94.17) .. (205,99) .. controls (205,103.83) and (200.52,107.75) .. (195,107.75) .. controls (189.48,107.75) and (185,103.83) .. (185,99) -- cycle ;
\draw    (407,97.5) -- (471,98) ;
\draw    (487,104.75) -- (531,142.5) ;
\draw    (488,90.5) -- (534,49.5) ;
\draw  [fill={rgb, 255:red, 0; green, 0; blue, 0 }  ,fill opacity=1 ] (471,98) .. controls (471,93.17) and (475.48,89.25) .. (481,89.25) .. controls (486.52,89.25) and (491,93.17) .. (491,98) .. controls (491,102.83) and (486.52,106.75) .. (481,106.75) .. controls (475.48,106.75) and (471,102.83) .. (471,98) -- cycle ;

\draw (101,86.4) node [anchor=north west][inner sep=0.75pt]    {$h_{1}$};
\draw (251,30.4) node [anchor=north west][inner sep=0.75pt]    {$h_{2}$};
\draw (247,146.9) node [anchor=north west][inner sep=0.75pt]    {$h_{3}$};
\draw (388,87.4) node [anchor=north west][inner sep=0.75pt]    {$h_{1}$};
\draw (537,31.4) node [anchor=north west][inner sep=0.75pt]    {$h_{2}$};
\draw (533,145.9) node [anchor=north west][inner sep=0.75pt]    {$h_{3}$};
\draw (153,155.4) node [anchor=north west][inner sep=0.75pt]    {$( i)$};
\draw (462,160.4) node [anchor=north west][inner sep=0.75pt]    {$( ii)$};
\end{tikzpicture}

  \caption{\textbf{(i)} helicity configuration $H = h_1 + h_2 + h_3 > 0$; \textbf{(ii)} helicity confirguration  $H = h_1 + h_2 + h_3 < 0$}
    \label{BFCWH}
\end{figure}
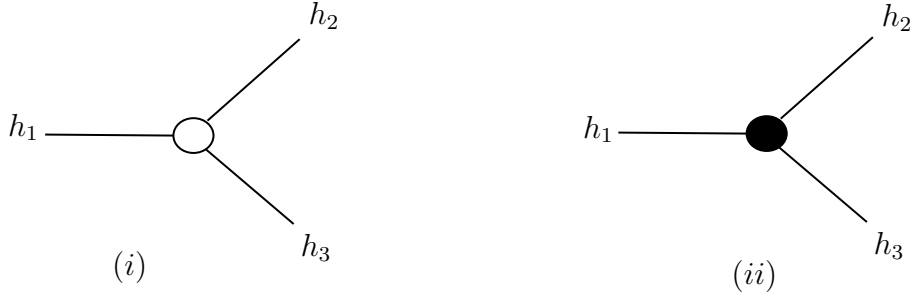

Under the action of little group, spinors transform as: $\chi_i \to t_i \chi_i$, and $\tilde{\chi}_i \to t^{-1}_i \tilde{\chi}_i$, while  $\epsilon^+_{\alpha \dot{\alpha}} \to t^{-2} \epsilon^+_{\alpha \dot{\alpha}}$
and  $\epsilon^-_{\alpha \dot{\alpha}} \to t^2 \epsilon^-_{\alpha \dot{\alpha}}$, whereas the momentum stays invariant. An amplitude consists of external lines which depends on the momenta of the particles, transforms under little group by a change of scale and follows the same transformation law as the spinors, and scales homogeneously for each and every particle. This type of scaling behavior induces the following master formula for the $3-$point amplitude in terms of their helicities $(h_1,h_2,h_3)$: 
\begin{equation}\label{Master formula}
\mathcal{M} ( 1^{h_1}, 2^{h_2},3^{h_3} ) = \left\lbrace 
\begin{array}{ll}
 c_{123} \, \langle 12\rangle^{h_3-h_2-h_1}  \langle 13 \rangle^{h_2-h_1-h_3}   \langle  23 \rangle^{h_1-h_2-h_3}, 
 \hspace{2mm} & h_1+h_2+h_3 < 0   \\
\tilde{c}_{123} \, [ 12 ]^{h_1+h_2-h_3}  [13 ]^{h_1+h_3-h_2}   [ 23 ]^{h_2+h_3-h_1}, 
\hspace{2mm} & h_1+h_2+h_3 > 0
\end{array}
\right.
\end{equation}
where $c_{123}$, $\tilde{c}_{123}$ are constants to be determined. $\langle ij\rangle$ and $[ij]$  denote the spinor products for particles with momenta $i,j$. Here we will follow same notation as in \cite{Arkani-Hamed:2012zlh} to denote the two configuration cases of helicities as shown in the figure below (\ref{BFCWH}).

 Let us now utilise \eqref{Master formula} to write down  the amplitude for the 3-point interactions between gluons and gluinos and graviton-gravitinos. Since we are interested in KLT relations -- calculations related to $(\tilde{g}\tilde{g}g)$ can be used to prove the equivalence between supergravity and Super Yang-Mills amplitude as stated in the introduction.  


\begin{table}[H]
\centering
\begin{tabular}{|c|p{1cm} p{1cm} p{1cm} p{1.3cm}|c|}
\hline
\cellcolor[gray]{0.94}\textbf{Interactions} & 
\multicolumn{4}{c|}{\cellcolor[gray]{0.94} \textbf{Input parameters}}
 & \cellcolor[gray]{0.94}\textbf{3-point amplitude}   \\ 
\cellcolor[gray]{0.94} & \cellcolor[gray]{0.94}$h_1$ &  \cellcolor[gray]{0.94}$h_2$ & \cellcolor[gray]{0.94}$h_3$ & \cellcolor[gray]{0.94}$H$ & \cellcolor[gray]{0.94} $\mathcal{M} ( 1^{h_1}, 2^{h_2},3^{h_3} ) $\\
\hline
\multirow{2}{*}{gluino-gluino-gluon $(\tilde{g}\tilde{g}g)$}  & $+\frac{1}{2}$ & $+1$ & $-\frac{1}{2}$  & $+1>0$ & $\widetilde{c}_{123} \frac{[12]^2}{[31]}$ \\ 
 & $-\frac{1}{2}$ & $-1$ &$+\frac{1}{2}$ & $-1<0$ & $c_{123} \frac{ \langle 12 \rangle^2}{ \langle 31 \rangle}$\\ 
 \hline
\multirow{ 2}{*}{graviton-graviton-graviton $(hhh)$} & $+2$ & $+2$ &$-2$ & $+2>0$ & $\widetilde{G}_{123} \frac{[12]^6}{ [23]^2 [31]^2}$\\ 
 & $-2$ & $-2$ &$+2$ & $-2<0$ & $G_{123} \frac{  \langle 12 \rangle^6}{ \langle 23 \rangle^2 \langle  31 \rangle^2}$\\ \hline
\multirow{ 2}{*}{gravitino-graviton-gravitino $(\tilde{h}h\tilde{h})$} & $+\frac{3}{2}$ & $+2$ &$-\frac{3}{2}$ & $+2>0$ & $\widetilde{g}_{123} \frac{[12]^5}{ [23] [31]^2}$\\ 
&  $-\frac{3}{2}$ & $-2$ &$+\frac{3}{2}$ & $-2<0$ & $g_{123} \frac{ \langle 12 \rangle^5}{ \langle 23 \rangle \langle 31 \rangle^2}$\\  \hline
\end{tabular}
\caption{3-point amplitudes for different interactions using \eqref{Master formula} is presented. The notation $H$ means $h_1+h_2+h_3$.}
\label{table2}
\end{table}
The amplitudes given in the second and third rows are the only ones that satisfy holomorphic and anti-holomorphic kinematic configurations for $\mN=1$ SUGRA three particle amplitudes 
 \cite{Cheung:2017pzi}.


\section{BCFW and the 4-point interactions in N=1 SUGRA}
\label{section3}

In this section, we aim to use a powerful tree level technique known as BCFW recursion relations \cite{Britto:2005fq,Britto:2004ap} which allows us to compute higher point partial amplitudes from lower point ones in a recursive way. Here we are going to utilise the constructive approach introduced in section \ref{section2} to write down the 4-point partial supergravity amplitudes. For our discussions, we will closely follow \cite{Schwartz:2014sze}. The BCFW transformations are
\begin{equation}
 \vert \hat{i} \rangle = \vert i \rangle, \quad [ \hat{i} \vert = [ i \vert + z [j \vert, 
\end{equation}
\begin{equation}
 [ \hat{j} \vert = [ j \vert, \quad \vert \hat{j} \rangle - z \vert i \rangle,
\end{equation}
with $z$ being a complex number, and the hat $\hat{i}$ denotes the shifted momenta of the particles $i$ and $j$.  
 Our convention is such that the particle $i$ resides on the left  and $j$  on the right side.  The shifted momentum formula of the particle related to the propagator which is on-shell,  is given by:
\begin{equation}\label{Pz}
    \hat{P}(z) = \vert \hat{P} \rangle [ \hat{P} \vert = \sum_{k =a}^b \vert k \rangle [ k \vert - z \vert i \rangle [ j \vert, 
\end{equation}
where the sum here is over particles $(a,b)$ on the right-hand side, so the pole is obtained by
\begin{equation}\label{pole}
    z^*_{a,b} = \frac{(\sum_{k =a}^b k)^2}{\sum_{k=a}^b \langle ik \rangle [kj]}.
\end{equation}
The BCFW recursion formula that we use in this paper is the following
\begin{align}\label{BCFWformula}
    \mathcal{M}(1 \dots n)  =\,& \sum_{a,b,h} \mathcal{M}(1, \dots , a-1, b+1, \dots , n \rightarrow \hat{P}^h) \nonumber\\ 
   \, & \times \frac{1}{(p_a + \dots p_b)^2} \mathcal{M}(\hat{P}^{-h} \rightarrow a, \dots , b),
\end{align}
where the right hand side are matrix elements evaluated in $z = z^*_{a,b}$  \cite{Schwartz:2014sze}.
 For our purpose, we compute the  three 4-point partial amplitudes associated  to SUGRA Compton scattering, showed in figure (\ref{Feynman Diagrams}) in section (\ref{section5}), using the program of BCFW recursion relations. We also work out the partial amplitudes using the helicity configuration $\mathcal{\Tilde{M}} (1^-_{\Tilde{h}}, 2^-_h, 3^+_{\Tilde{h}}, 4^+_h)$ for all BCFW diagrams.
 
\subsection{BCFW Diagram A}
 In this case we have two BCFW diagrams associated to $s_{14}$ channel and Feynman diagram A of figure (\ref{Feynman Diagrams}), which are shown in the figure below:  
\begin{figure}[H]
    \centering
\tikzset{every picture/.style={line width=0.75pt}} 

\begin{tikzpicture}[x=0.75pt,y=0.75pt,yscale=-1,xscale=1]

\draw    (45,38) -- (92,76) ;
\draw  [fill={rgb, 255:red, 0; green, 0; blue, 0 }  ,fill opacity=1 ] (90,83) .. controls (90,77.48) and (94.7,73) .. (100.5,73) .. controls (106.3,73) and (111,77.48) .. (111,83) .. controls (111,88.52) and (106.3,93) .. (100.5,93) .. controls (94.7,93) and (90,88.52) .. (90,83) -- cycle ;
\draw    (43,119) -- (93,91) ;
\draw    (111,83) -- (191,83) ;
\draw   (191,83) .. controls (191,77.48) and (195.7,73) .. (201.5,73) .. controls (207.3,73) and (212,77.48) .. (212,83) .. controls (212,88.52) and (207.3,93) .. (201.5,93) .. controls (195.7,93) and (191,88.52) .. (191,83) -- cycle ;
\draw    (208,75) -- (259,40) ;
\draw    (207,91) -- (258,127) ;
\draw    (387,41) -- (434,79) ;
\draw   (432,86) .. controls (432,80.48) and (436.7,76) .. (442.5,76) .. controls (448.3,76) and (453,80.48) .. (453,86) .. controls (453,91.52) and (448.3,96) .. (442.5,96) .. controls (436.7,96) and (432,91.52) .. (432,86) -- cycle ;
\draw    (385,122) -- (435,94) ;
\draw    (453,86) -- (533,86) ;
\draw  [fill={rgb, 255:red, 0; green, 0; blue, 0 }  ,fill opacity=1 ] (533,86) .. controls (533,80.48) and (537.7,76) .. (543.5,76) .. controls (549.3,76) and (554,80.48) .. (554,86) .. controls (554,91.52) and (549.3,96) .. (543.5,96) .. controls (537.7,96) and (533,91.52) .. (533,86) -- cycle ;
\draw    (550,78) -- (601,43) ;
\draw    (550,95) -- (601,130) ;

\draw (258,123.4) node [anchor=north west][inner sep=0.75pt]    {$2{^{-}}$};
\draw (22,120.4) node [anchor=north west][inner sep=0.75pt]    {$\hat{1}^{-}$};
\draw (23,22.4) node [anchor=north west][inner sep=0.75pt]    {$4^{+}$};
\draw (264,14.4) node [anchor=north west][inner sep=0.75pt]    {$\hat{3}^{+}$};
\draw (599,131.4) node [anchor=north west][inner sep=0.75pt]    {$2^{-}$};
\draw (363,123.4) node [anchor=north west][inner sep=0.75pt]    {$\hat{1}^{-}$};
\draw (365,24.4) node [anchor=north west][inner sep=0.75pt]    {$4^{+}$};
\draw (605,19.4) node [anchor=north west][inner sep=0.75pt]    {$\hat{3}^{+}$};
\draw (115,63.4) node [anchor=north west][inner sep=0.75pt]    {$-$};
\draw (176,62.4) node [anchor=north west][inner sep=0.75pt]    {$+$};
\draw (455,66.4) node [anchor=north west][inner sep=0.75pt]    {$+$};
\draw (519,66.4) node [anchor=north west][inner sep=0.75pt]    {$-$};
\draw (135,127.4) node [anchor=north west][inner sep=0.75pt]    {$( a)$};
\draw (495,128.4) node [anchor=north west][inner sep=0.75pt]    {$( b)$};

\end{tikzpicture}

  \caption{BCFW Diagram A}
    \label{BCFWdiagramaA}
\end{figure}
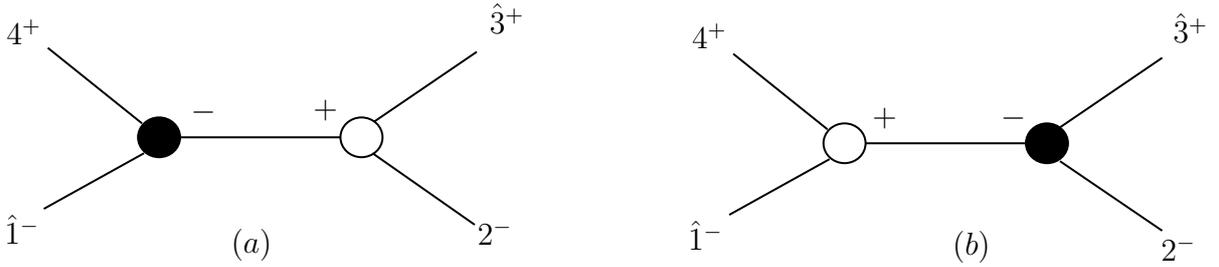
The  helicity amplitude configuration which follows from equation (\ref{BCFWformula}) is
\begin{align}\label{BCFWA}
\mathcal{\widetilde{M}}_A(1^{-\frac{3}{2}}_{\widetilde{h}}, 4^{+2}_h, 3^{+\frac{3}{2}}_{\widetilde{h}}, 2^{-2}_h)  = \,& \mathcal{\widetilde{M}}_a(1^{-\frac{3}{2}}, 4^{+2}, \hat{P}^{-\frac{3}{2}}) \frac{1}{s_{14}} \mathcal{\widetilde{M}}_a(-\hat{P}^{+\frac{3}{2}}, 3^{+ \frac{3}{2}}, 2^{-2}) \nonumber\\ 
&  + \mathcal{\widetilde{M}}_b(1^{- \frac{3}{2}}, 4^{+2}, \hat{P}^{+ \frac{3}{2}})   \frac{1}{s_{14}} \mathcal{\widetilde{M}}_b(- \hat{P}^{- \frac{3}{2}}, 3^{+ \frac{3}{2}}, 2^{-2}).
\end{align}
Taking the shift as $i = 1$ and $j=3$, which is a $(-,+)$ combination then is a well behaved choice as $z \to \infty$. Then the BCFW transformations become
\begin{equation}
    \vert \hat{1} \rangle = \vert 1 \rangle, \quad [ \hat{1} \vert = [ 1 \vert +  z^*_{2,3} [3 \vert,
\end{equation}
\begin{equation}
    [ \hat{3} \vert = [ 3 \vert, \quad \vert \hat{3} \rangle = \vert 3 \rangle - z^*_{2,3} \vert 1 \rangle.
\end{equation}
Using equation (\ref{pole}) the pole of the channel $s_{14}$ is
\begin{equation}
    z^*_{2,3} = \frac{(p_2 + p_3)^2}{ \langle 12 \rangle [23] + \langle 13 \rangle [33]} = \frac{\langle 23 \rangle [32]}{\langle 12 \rangle [23]} = - \frac{\langle 23 \rangle}{ \langle 12 \rangle}.
\end{equation}
According to  table (\ref{table2}), in (\ref{BCFWA}), the terms $\mathcal{\widetilde{M}}(1^{-\frac{3}{2}}, 4^{+2}, \hat{P}^{-\frac{3}{2}})$ and $\mathcal{\widetilde{M}}(-\hat{P}^{+\frac{3}{2}}, 3^{+ \frac{3}{2}}, 2^{-2})$ must vanish, so we are only left with contributions coming from amplitude $b$ 
\begin{align}\label{Mb}
 \mathcal{\widetilde{M}}_A( 1^{-\frac{3}{2}}_{\widetilde{h}}, 4^{+2}_h, 3^{+\frac{3}{2}}_{\widetilde{h}}, 2^{-2}_h) &= \mathcal{\widetilde{M}}_b(\hat{1}^{- \frac{3}{2}}, 4^{+2}, \hat{P}^{+ \frac{3}{2}})   \frac{1}{s_{14}} \mathcal{\widetilde{M}}_b(- \hat{P}^{- \frac{3}{2}}, \hat{3}^{+ \frac{3}{2}}, 2^{-2}) \\ \nonumber
 &= \frac{[\hat{P} 4]^5}{[4 \hat{1}] [\hat{1} \hat{P}]^2} \frac{1}{ \langle 14 \rangle [41]} \frac{ \langle (- \hat{P}) 2 \rangle^5}{ \langle 2 \hat{3} \rangle \langle \hat{3} (- \hat{P}) \rangle^2}.
\end{align}
where we have used previous results coming from the third column of the  table (\ref{table2})-- suppressing factors $\widetilde{g}_{123}$ and $g_{123}$. At three point amplitude level, spinors with momenta $-P$ and $P$ are related as
\begin{equation}\label{minusPP}
    \vert (-P) \rangle = i \vert P \rangle, \quad \vert (- P) ] = i \vert P ], \quad
    \langle (-P) \vert = i \langle P \vert, \quad [ (-P) \vert = i [ P \vert.
\end{equation}
 In this approach the shifted momentum of (\ref{Pz}) takes the following form
\begin{equation}\label{minusPP1}
  \vert \hat{P} \rangle [ \hat{P} \vert = \vert 2 \rangle [2 \vert + \vert 3 \rangle [ 3 \vert - z^*_{2,3} \vert 1 \rangle [ 3 \vert ,
\end{equation}
Finally, applying both \eqref{minusPP} and \eqref{minusPP1}, the numerator  of \eqref{Mb} becomes 
\begin{align}
    \langle 2 \hat{P} \rangle [\hat{P} 4] & = \langle 23 \rangle [34] + \frac{\langle 23 \rangle}{\langle 12 \rangle} \langle 21 \rangle [34] \nonumber\\ 
    &= \langle 23 \rangle [34] - \langle 23 \rangle [34]. 
\end{align}
Therefore, the amplitude of \eqref{Mb} vanishes 
\begin{equation}\label{MBCFWA}
  \mathcal{\widetilde{M}}_A(1^{-\frac{3}{2}}_{\widetilde{h}}, 4^{+2}_h, 3^{+\frac{3}{2}}_{\widetilde{h}}, 2^{-2}_h) = 0   
\end{equation}
\subsection{BCFW Diagram B}
We have two BCFW diagrams associated to $s_{12}$ channel as shown in Figure (\ref{BCFWdiagramB}):
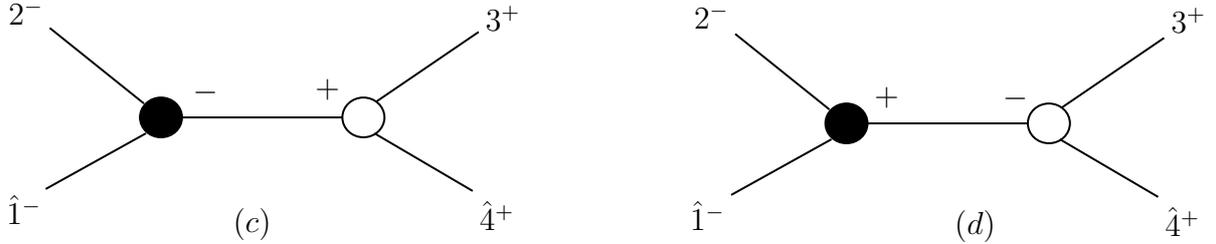
\begin{figure}[H]
    \centering
\tikzset{every picture/.style={line width=0.75pt}} 

\begin{tikzpicture}[x=0.75pt,y=0.75pt,yscale=-1,xscale=1]

\draw    (45,38) -- (92,76) ;
\draw  [fill={rgb, 255:red, 0; green, 0; blue, 0 }  ,fill opacity=1 ] (90,83) .. controls (90,77.48) and (94.7,73) .. (100.5,73) .. controls (106.3,73) and (111,77.48) .. (111,83) .. controls (111,88.52) and (106.3,93) .. (100.5,93) .. controls (94.7,93) and (90,88.52) .. (90,83) -- cycle ;
\draw    (43,119) -- (93,91) ;
\draw    (111,83) -- (191,83) ;
\draw   (191,83) .. controls (191,77.48) and (195.7,73) .. (201.5,73) .. controls (207.3,73) and (212,77.48) .. (212,83) .. controls (212,88.52) and (207.3,93) .. (201.5,93) .. controls (195.7,93) and (191,88.52) .. (191,83) -- cycle ;
\draw    (208,75) -- (259,40) ;
\draw    (207,91) -- (256,120) ;
\draw    (387,41) -- (434,79) ;
\draw  [fill={rgb, 255:red, 0; green, 0; blue, 0 }  ,fill opacity=1 ] (432,86) .. controls (432,80.48) and (436.7,76) .. (442.5,76) .. controls (448.3,76) and (453,80.48) .. (453,86) .. controls (453,91.52) and (448.3,96) .. (442.5,96) .. controls (436.7,96) and (432,91.52) .. (432,86) -- cycle ;
\draw    (385,122) -- (435,94) ;
\draw    (453,86) -- (533,86) ;
\draw   (533,86) .. controls (533,80.48) and (537.7,76) .. (543.5,76) .. controls (549.3,76) and (554,80.48) .. (554,86) .. controls (554,91.52) and (549.3,96) .. (543.5,96) .. controls (537.7,96) and (533,91.52) .. (533,86) -- cycle ;
\draw    (550,78) -- (601,43) ;
\draw    (549,94) -- (598,123) ;

\draw (258,123.4) node [anchor=north west][inner sep=0.75pt]    {$\hat{4}^{+}$};
\draw (22,120.4) node [anchor=north west][inner sep=0.75pt]    {$\hat{1}^{-}$};
\draw (23,22.4) node [anchor=north west][inner sep=0.75pt]    {$2^{-}$};
\draw (261,25.4) node [anchor=north west][inner sep=0.75pt]    {$3^{+}$};
\draw (600,126.4) node [anchor=north west][inner sep=0.75pt]    {$\hat{4}^{+}$};
\draw (363,123.4) node [anchor=north west][inner sep=0.75pt]    {$\hat{1}^{-}$};
\draw (365,24.4) node [anchor=north west][inner sep=0.75pt]    {$2^{-}$};
\draw (603,26.4) node [anchor=north west][inner sep=0.75pt]    {$3^{+}$};
\draw (115,63.4) node [anchor=north west][inner sep=0.75pt]    {$-$};
\draw (176,62.4) node [anchor=north west][inner sep=0.75pt]    {$+$};
\draw (455,66.4) node [anchor=north west][inner sep=0.75pt]    {$+$};
\draw (519,66.4) node [anchor=north west][inner sep=0.75pt]    {$-$};
\draw (135,127.4) node [anchor=north west][inner sep=0.75pt]    {$( c)$};
\draw (495,128.4) node [anchor=north west][inner sep=0.75pt]    {$( d)$};
\end{tikzpicture}
\caption{BCFW Diagram B}
    \label{BCFWdiagramB}
\end{figure}

The corresponding amplitude configuration is the following
\begin{align}
    \widetilde{\mathcal{M}}_B(1^{- \frac{3}{2}},  2^{-2}, 3^{+ \frac{3}{2}}, 4^{+2} ) =\,& \widetilde{\mathcal{M}}_c ( 1^{- \frac{3}{2}}, 2^{-2}, \hat{P}^{- \frac{3}{2}}) \frac{1}{s_{12}} \widetilde{\mathcal{M}}_c( -\hat{P}^{+\frac{3}{2}}, 3^{+ \frac{3}{2}}, 4^{+2}) \\ \nonumber
    & + \widetilde{\mathcal{M}}_d(1^{- \frac{3}{2}}, 2^{-2}, \hat{P}^{+ \frac{3}{2}}) \frac{1}{s_{12}} \widetilde{\mathcal{M}}_d(- \hat{P}^{- \frac{3}{2}}, 3^{+ \frac{3}{2}}, 4^{+2}).
\end{align}
In this case, we consider  $i= 1$ and $j=4$, again using a $(- ,+)$ combination,  which is a well behaved shift. Then BCFW transformations become
\begin{equation}
    \vert \hat{1} \rangle = \vert 1 \rangle, \quad [ \hat{1} \vert = [1 \vert + z^*_{3,4} [ 4 \vert,
\end{equation}
\begin{equation}
    [\hat{4} \vert = [4 \vert, \quad \vert \hat{4} \rangle = \vert 4 \rangle - z^*_{3,4} \vert 1 \rangle.
\end{equation}
the pole in this case is
\begin{equation}
    z^*_{3,4} = \frac{\langle 34 \rangle [43]}{ \langle 13 \rangle [34]} = - \frac{\langle 34 \rangle}{ \langle 13 \rangle}.
\end{equation}
The same procedure that we developed above in the partial amplitude for BCFW diagram A, are also valid here, hence we only need to calculate the contribution from partial amplitude $d$
\begin{align}\label{Md}
  \widetilde{\mathcal{M}}_B(1^{- \frac{3}{2}},  2^{-2}, 3^{+ \frac{3}{2}}, 4^{+2} ) &= \widetilde{\mathcal{M}}_d(\hat{1}^{- \frac{3}{2}}, 2^{-2}, \hat{P}^{+ \frac{3}{2}}) \frac{1}{s_{12}} \widetilde{\mathcal{M}}_d(- \hat{P}^{- \frac{3}{2}}, 3^{+ \frac{3}{2}}, \hat{4}^{+2})\nonumber \\ 
  & = \frac{ \langle \hat{1}2 \rangle^5}{ \langle 2 \hat{P} \rangle  \langle \hat{P} \hat{1} \rangle^2} \frac{1}{ \langle 34 \rangle [43]} \frac{[3 \hat{4}]^5}{ [ \hat{4} (- \hat{P})] [ (-\hat{P}) 3]^2} \nonumber\\ 
  & = i \frac{ \langle 12 \rangle^5 [34]^5}{ \langle 2 \hat{P} \rangle [ \hat{P} 4] \langle 34 \rangle [43] \langle 1 \hat{P} \rangle^2 [\hat{P} 3]^2},
\end{align}
where again we suppressed factors $\tilde{g}_{123}$ and $g_{123}$ using results of third column of table (\ref{table2}). The momentum of the internal line is
\begin{equation}
    \vert \hat{P} \rangle [ \hat{P} \vert = \vert 3 \rangle [3 \vert + \vert 4 \rangle [ 4 \vert + \frac{\langle 34 \rangle}{ \langle 13 \rangle} \vert 1 \rangle [ 4 \vert.
\end{equation}
Therefore, the denominator of \eqref{Md}  becomes
\begin{align}
    \langle 2 \hat{P} \rangle [\hat{P} 4] =\,& \langle 23 \rangle [34],\\
    \langle 1 \hat{P} \rangle [\hat{P} 3] = \,&\langle 14 \rangle [43].
\end{align}
The final form of the amplitude becomes
\begin{equation}\label{MBCFWB}
     \widetilde{\mathcal{M}}_B(1^{- \frac{3}{2}},  2^{-2}, 3^{+ \frac{3}{2}}, 4^{+2} ) = -i \frac{\langle 12 \rangle^6 [34]}{ \langle 12 \rangle \langle 23 \rangle \langle 34 \rangle  \langle 41 \rangle^2}.
\end{equation}

\subsection{BCFW Diagram C}
We have two BCFW diagrams associated to $s_{13}$ channel  shown in the  figure below and Feynman diagram C of (\ref{Feynman Diagrams}):
\begin{figure}[H]
    \centering
\tikzset{every picture/.style={line width=0.75pt}} 

\begin{tikzpicture}[x=0.75pt,y=0.75pt,yscale=-1,xscale=1]

\draw    (45,38) -- (92,76) ;
\draw  [fill={rgb, 255:red, 0; green, 0; blue, 0 }  ,fill opacity=1 ] (90,83) .. controls (90,77.48) and (94.7,73) .. (100.5,73) .. controls (106.3,73) and (111,77.48) .. (111,83) .. controls (111,88.52) and (106.3,93) .. (100.5,93) .. controls (94.7,93) and (90,88.52) .. (90,83) -- cycle ;
\draw    (43,119) -- (93,91) ;
\draw    (111,83) -- (191,83) ;
\draw   (191,83) .. controls (191,77.48) and (195.7,73) .. (201.5,73) .. controls (207.3,73) and (212,77.48) .. (212,83) .. controls (212,88.52) and (207.3,93) .. (201.5,93) .. controls (195.7,93) and (191,88.52) .. (191,83) -- cycle ;
\draw    (208,75) -- (259,40) ;
\draw    (207,91) -- (256,120) ;
\draw    (387,41) -- (434,79) ;
\draw   (432,86) .. controls (432,80.48) and (436.7,76) .. (442.5,76) .. controls (448.3,76) and (453,80.48) .. (453,86) .. controls (453,91.52) and (448.3,96) .. (442.5,96) .. controls (436.7,96) and (432,91.52) .. (432,86) -- cycle ;
\draw    (385,122) -- (435,94) ;
\draw    (453,86) -- (533,86) ;
\draw  [fill={rgb, 255:red, 0; green, 0; blue, 0 }  ,fill opacity=1 ] (533,86) .. controls (533,80.48) and (537.7,76) .. (543.5,76) .. controls (549.3,76) and (554,80.48) .. (554,86) .. controls (554,91.52) and (549.3,96) .. (543.5,96) .. controls (537.7,96) and (533,91.52) .. (533,86) -- cycle ;
\draw    (550,78) -- (601,43) ;
\draw    (549,94) -- (598,123) ;

\draw (258,123.4) node [anchor=north west][inner sep=0.75pt]    {$\hat{4}^{+}$};
\draw (22,120.4) node [anchor=north west][inner sep=0.75pt]    {$\hat{1}^{-}$};
\draw (23,22.4) node [anchor=north west][inner sep=0.75pt]    {$3^{+}$};
\draw (261,25.4) node [anchor=north west][inner sep=0.75pt]    {$2^{-}$};
\draw (600,126.4) node [anchor=north west][inner sep=0.75pt]    {$\hat{4}^{+}$};
\draw (363,123.4) node [anchor=north west][inner sep=0.75pt]    {$\hat{1}^{-}$};
\draw (365,24.4) node [anchor=north west][inner sep=0.75pt]    {$3^{+}$};
\draw (603,26.4) node [anchor=north west][inner sep=0.75pt]    {$2^{-}$};
\draw (115,63.4) node [anchor=north west][inner sep=0.75pt]    {$-$};
\draw (176,62.4) node [anchor=north west][inner sep=0.75pt]    {$+$};
\draw (455,66.4) node [anchor=north west][inner sep=0.75pt]    {$+$};
\draw (519,66.4) node [anchor=north west][inner sep=0.75pt]    {$-$};
\draw (135,127.4) node [anchor=north west][inner sep=0.75pt]    {$( e)$};
\draw (495,128.4) node [anchor=north west][inner sep=0.75pt]    {$( f)$};
\end{tikzpicture}
\caption{BCFW Diagram C}
\label{BCFWdiagramC}
\end{figure}
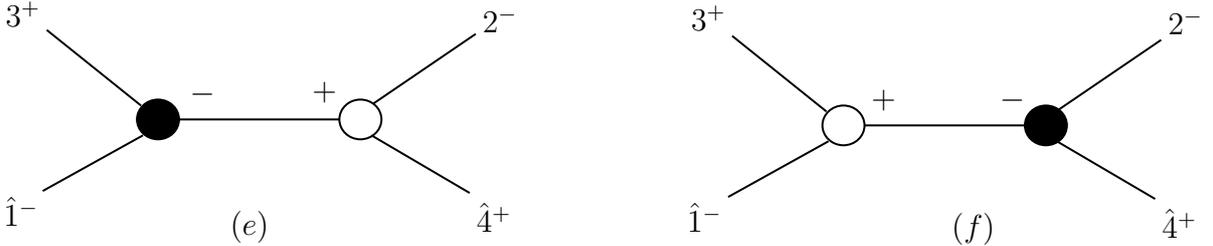

The amplitude configurations for this diagram is
\begin{align}
    \widetilde{\mathcal{M}}_C(1^{- \frac{3}{2}}, 3^{+ \frac{3}{2}}, 2^{-2}, 4^{+2}) =\,& \widetilde{\mathcal{M}}_e( 1^{- \frac{3}{2}}, 3^{+ \frac{3}{2}}, \hat{P}^{-2}) \frac{i}{s_{13}} \widetilde{\mathcal{M}}_e(- \hat{P}^{+2}, 2^{-2}, 4^{+2}) \\ \nonumber
    & + \widetilde{\mathcal{M}}_f(1^{- \frac{3}{2}}, 3^{+ \frac{3}{2}}, \hat{P}^{+2}) \frac{i}{s_{13}} \widetilde{\mathcal{M}}_f(- \hat{P}^{-2}, 2^{-2}, 4^{+2}).
\end{align}
Here we choose the shift as $i =1$ and $j= 4$, again a $(-,+)$ combination. The BCFW transformations here are
\begin{equation}
    \vert \hat{1} \rangle = \vert 1 \rangle, \quad [ \hat{1} \vert = [1 \vert + z^*_{2,4} [4 \vert,
\end{equation}
\begin{equation}
    [ \hat{4} \vert = [4 \vert, \quad \vert \hat{4} \rangle = \vert 4 \rangle - z^*_{2,4} \vert 1 \rangle.
\end{equation}
Now, the pole is
\begin{equation}
z^*_{2,4} = \frac{\langle 24 \rangle [42]}{\langle 12 \rangle [24]} = - \frac{\langle 24 \rangle}{ \langle 12 \rangle}.
\end{equation}
In this case, both of the terms are non-vanishing
\begin{align}\label{MC}
   \widetilde{\mathcal{M}}_C(1^{- \frac{3}{2}}, 3^{+ \frac{3}{2}}, 2^{-2}, 4^{+2}) =   \frac{ \langle  \hat{1} \hat{P} \rangle^5}{ \langle \hat{P} 3 \rangle \langle 3 \hat{1} \rangle^2} \frac{i}{s_{13}} \frac{[ (- \hat{P}) \hat{4}]^6}{[\hat{4}2]^2 [2 (- \hat{P} )]^2} + \frac{[3 \hat{P}]^5}{[\hat{P} \hat{1}] [\hat{1}3]^2} \frac{i}{s_{13}} \frac{\langle (- \hat{P}) 2 \rangle^6}{ \langle 2 \hat{4} \rangle^2 \langle \hat{4} (- \hat{P}) \rangle^2}.
\end{align}
On the other hand, the momentum of the internal line is
\begin{equation}
    \vert \hat{P} \rangle [ \hat{P} \vert = \vert 2 \rangle [2 \vert + \vert 4 \rangle [4 \vert + \frac{\langle 24 \rangle}{ \langle 12 \rangle} \vert 1 \rangle [ 4 \vert,
\end{equation}
for the second term of (\ref{MC}), we have
\begin{equation}
    \langle 2 \hat{P} \rangle [\hat{P} 3] = \langle 24 \rangle [43] + \frac{\langle 24 \rangle}{\langle 12 \rangle} \langle 21 \rangle [43] = 0,
\end{equation}
therefore we see that  only  the first term  of (\ref{MC}) survives, related to diagram (e) of \ref{BCFWdiagramC}, so 
 according to the results of third row from table (\ref{table2}), the amplitude is
\begin{align}\label{Mc13}
  \widetilde{\mathcal{M}}_C(1^{- \frac{3}{2}}, 3^{+ \frac{3}{2}}, 2^{-2}, 4^{+2}) &=  i \frac{ \langle 1 \hat{P} \rangle^5 [\hat{P} 4]^5 [\hat{P} 4]}{\langle 31 \rangle^2 [42]^2 \langle 3 \hat{P} \rangle [\hat{P} 2] [\hat{P} 2] \langle 24 \rangle [42]},
\end{align}
then, we find that:  
\begin{equation}
    \langle 1 \hat{P} \rangle [\hat{P} 4] = \langle 12 \rangle [24] = - \langle 13 \rangle [34],
\end{equation}
\begin{equation}
    \langle 3 \hat{P} \rangle [\hat{P} 2] = \langle 34 \rangle [42] + \frac{\langle 24 \rangle}{\langle 12 \rangle} \langle 31 \rangle [42],
\end{equation}
\begin{equation}
    \frac{[\hat{P} 4]}{[\hat{P} 2]} = \frac{\langle 1 \hat{P} \rangle [\hat{P} 4]}{\langle 1 \hat{P} \rangle [\hat{P} 2]} = \frac{\langle 12 \rangle [24]}{ \langle 14 \rangle [42]} = - \frac{\langle 12 \rangle}{ \langle 14 \rangle}.
\end{equation}
Substituting above results in \eqref{Mc13}, the final amplitude form is
\begin{equation}\label{MBCFWC}
   \widetilde{\mathcal{M}}_C(1^{- \frac{3}{2}}, 3^{+ \frac{3}{2}}, 2^{-2}, 4^{+2} )  =  i \frac{\langle 12 \rangle^6 [34]}{ \langle 13 \rangle  \langle 32 \rangle \langle 24 \rangle \langle 41 \rangle^2}.
\end{equation}
Hence, we have two partial amplitudes contributions from channels $s_{12}$ and $s_{13}$. These are 4-points interactions between gravitons and gravitinos. 

\section{KLT relations and the action of N=1 SUGRA} \label{section4}

In this section we  will derive the form of KLT relations such that we can compare with results obtained from the master equation following the constructive approach in section \eqref{section2}. We shall take \eqref{VhHh3} and \eqref{VHHH3} that comes from  \cite{Bjerrum-Bohr:2010eeh}, then contracting them with polarization tensors \eqref{h1}-\eqref{h11} to derive expressions for the 3-point interactions $hhh$ and $h\tilde{h}\tilde{h}$ in terms of the SUSY-QCD interactions  $ggg$ and $g\tilde{g}\tilde{g}$.  Details of the $\mN=1$ SUGRA Lagrangian in terms of Weyl spinors are presented in \ref{appB}, including the invariance under gauge and global super symmetric transformations. \\
The $3-$vertex of two gravitinos  one graviton, which is the product of  a SYM vertex that involves two gluinos one gluon and a YM vertex with all gluons, as follows \cite{Bjerrum-Bohr:2010eeh}:
\begin{align}\label{VhHh3}
		V^{\mu, \alpha \beta, \nu}_{\widetilde{h} h \widetilde{h}} (p_1, p_2, p_3) &=  - V^{\alpha}_{\widetilde{g} g \widetilde{g}}(p_1, p_2, p_3) \times V^{\mu \beta \nu}_{ggg}(p_1, p_2, p_3) \\ \nonumber
  &= - i \left( \frac{i}{\sqrt{2}} \gamma^{\alpha} \right) \times \left( \frac{i}{\sqrt{2}}     [ (p_1 - p_2)^{\nu} \eta^{\beta \mu}  + (p_2 - p_3)^{\mu} \eta^{\beta \nu} + (p_3 - p_1)^{\beta} \eta^{\mu \nu} ]   \right).
\end{align}
For the three graviton vertex, which is the product of two YM vertices that involves only gluons, we have:
\begin{align}\label{VHHH3}
    V^{\mu \nu, \alpha \beta, \rho \sigma}_{hhh}(p_1,p_2,p_3) = \,&- i \left( \frac{i}{\sqrt{2}} \right)^2 (        (p_1 - p_2)^{\rho} \eta^{\mu \alpha} + (p_2 - p_3)^{\mu} \eta^{\alpha \rho} + (p_3 - p_1)^{\alpha} \eta^{\rho \mu}    ) \times \\ \nonumber
    & (  (p_1 - p_2)^{\sigma} \eta^{\beta \nu} + (p_2 - p_3)^{\nu} \eta^{\sigma \beta} + (p_3 - p_1)^{\beta} \eta^{\nu \sigma}        ).
\end{align}
The constraint $p_1 + p_2 + p_3 =0$ implies:
\begin{equation}\label{3momento}
  p_1- p_2 = 2p_1 + p_3, \quad p_2 - p_3 =  2p_2 + p_1, \quad p_3 - p_1 = 2p_3 + p_2,
\end{equation}
the relations \eqref{3momento} help us to rewrite conveniently the above vertices as follows:
\begin{align}
V^{\mu, \alpha \beta, \nu}_{\widetilde{h} h \widetilde{h}} (p_1, p_2, p_3) =  \,& - i \left( \frac{i}{\sqrt{2}} \gamma^{\alpha} \right)\nonumber \\
&\times \left( \frac{i}{\sqrt{2}}    [ (2p_1 + p_3)^{\nu} \eta^{\beta \mu}  + (2p_2 + p_1)^{\mu} \eta^{\beta \nu} + (2p_3 + p_2)^{\beta} \eta^{\mu \nu}   ]   \right), \label{Vh} \\
    V^{\mu \nu, \alpha \beta, \rho \sigma}_{hhh}(p_1,p_2,p_3) =\,& - i \left(\frac{i}{\sqrt{2}} \right)^2\nonumber\\
    &\times (        (2p_1 + p_3)^{\rho} \eta^{\mu \alpha} + (2p_2 + p_1)^{\mu} \eta^{\alpha \rho} + (2p_3 + p_2)^{\alpha} \eta^{\rho \mu}    ) \times \nonumber\\ 
    &\, (  (2p_1 + p_3)^{\sigma} \eta^{\beta \nu} + (2p_2 + p_1)^{\nu} \eta^{\sigma \beta} + (2p_3 + p_2)^{\beta} \eta^{\nu \sigma}        )\label{Vhhh}.
\end{align}
Now we are going to compute the three point interactions for $\Tilde{h} h \longrightarrow \Tilde{h} h $. We already know the three point vertices necessary to calculate the $hhh$ and $\Tilde{h} h \Tilde{h}$ amplitudes. We actually have only two possible helicity configurations that do not vanish as we show below. The three graviton amplitude $hhh$ with helicity configuration $h_1 = +2$, $h_2 = +2$ and $h_3 = -2$ is
 \begin{equation}
\widetilde{ \mathcal{M}}(1^{+2}, 2^{+2}, 3^{-2}) = V^{\mu \nu, \alpha \beta, \rho \sigma}_{hhh}(p_1, p_2, p_3) \times \epsilon^{++}_{h, \mu \nu}(p_1, q) \epsilon^{++}_{h, \alpha \beta}(p_2,l) \epsilon^{--}_{h, \rho \sigma}(p_3, r),     
 \end{equation}
with $q,l,r$ being respective reference momenta of $p_1, p_2, p_3$. Here we are contracting the three graviton vertex (\ref{Vhhh}) with polarization tensors  \eqref{h00}, \eqref{h11} and  using on-shell conditions (\ref{OSR}), then we have our three graviton amplitude in terms of gluon polarization vectors as
\begin{align}
\widetilde{ \mathcal{M}}(1^{+2}, 2^{+2}, 3^{-2}) =\,& -4i \left( \frac{i}{\sqrt{2}} \right)^2  
 ( (\epsilon^+_g(p_1,q) \cdot p_2) \times ( \epsilon^+_g(p_2,l) \cdot \epsilon^-_g(p_3,r)  ) \\ \nonumber
 & + (\epsilon^+_g(p_2,l) \cdot p_3) \times ( \epsilon^-_g(p_3,r) \cdot \epsilon^+_g(p_1,q)) )^2,   
\end{align}
rewriting all contractions of this equation  explicity in terms of Weyl helicity spinor using (\ref{epj0}) and (\ref{e1e0}), we obtain 
\begin{align}\label{M001}
(\epsilon^+_g(p_1,q) \cdot p_2) \times &( \epsilon^+_g(p_2,l) \cdot \epsilon^-_g(p_3,r)  ) + (\epsilon^+_g(p_2,l) \cdot p_3) \times ( \epsilon^-_g(p_3,r) \cdot \epsilon^+_g(p_1,q))   \nonumber\\ 
& = \frac{[12] \langle 2q \rangle }{ \sqrt{2} \langle q1 \rangle } \times \frac{ \langle 3l \rangle [2r]}{ [3r] \langle l2 \rangle }  \quad + \quad \frac{[23] \langle 3l \rangle}{ \sqrt{2}   \langle l2 \rangle} \times \frac{ \langle 3q \rangle [1r]}{[3r] \langle q1 \rangle }  \nonumber\\
& = \frac{ \langle 3l \rangle \langle 3q \rangle}{ \sqrt{2} \langle q1 \rangle \langle l2 \rangle [3r]  } \times \left(     [31] [2r] + [23][1r]      \right) \nonumber\\ 
&= \frac{ \langle 3l \rangle \langle 3q \rangle}{ \sqrt{2} \langle q1 \rangle \langle l2 \rangle [3r]  }  \times ( - [3r] [12]  )\nonumber \\ 
&= - \frac{ \langle l3 \rangle \langle q3 \rangle [12]}{ \sqrt{2} \langle l2 \rangle \langle q1 \rangle } \nonumber\\ 
&= - \frac{[12]^3}{  \sqrt{2} [23] [31]}
\end{align}
where we applied the following relation: $[12] \langle 2q \rangle = [ 1 \vert \not\! 2 \vert q \rangle = [ 1 \vert (- \not \!1 - \not \! 3) \vert q \rangle = - [ 1 \vert \not \! 3 \vert q \rangle = - [13] \langle 3q \rangle = [31] \langle 3q \rangle $ to arrive at \eqref{M001}. We have used the Schouten identity (\ref{Schouten}) for square brackets. Finally, we applied momentum conservation (\ref{MCrelation}) for $n=4$, therefore we have  the relation
\begin{equation}
    \langle l2 \rangle [21] + \langle l3 \rangle [31] = 0,
\end{equation}
which imply that
\begin{equation}
    \frac{\langle l3 \rangle}{ \langle l2 \rangle} = \frac{[12]}{ [31]}.
\end{equation}
Doing the same procedure for
\begin{equation}
    \langle q1 \rangle [12] + \langle q3 \rangle [32] = 0
\end{equation}
then
\begin{equation}
    \frac{\langle q3 \rangle}{ \langle q1 \rangle} = \frac{[12]}{ [23]}.
\end{equation}

The final result is
\begin{equation}\label{Gravitones0}
\widetilde{ \mathcal{M}}(1^{+2}, 2^{+2}, 3^{-2}) = -4i \left(\frac{i}{\sqrt{2}} \right)^2  \left( - \frac{[12]^3}{  \sqrt{2} [23] [31]} \right)^2    = i \frac{[12]^6}{[23]^2 [31]^2}.  
\end{equation}
We can observe that this 3-point amplitude does not depend on the choice of a reference momentum. The other amplitude in consideration is $ \mathcal{M}(1^{-2}, 2^{-2}, 3^{+2})$, in this case we have
\begin{equation}
\widetilde{ \mathcal{M}}(1^{-2}, 2^{-2}, 3^{+2}) = V^{\mu \nu, \alpha \beta, \rho \sigma}_{hhh}(p_1, p_2, p_3) \times \epsilon^{--}_{h, \mu \nu}(p_1, q) \epsilon^{--}_{h, \alpha \beta}(p_2,l) \epsilon^{++}_{h, \rho \sigma}(p_3, r),    
\end{equation}
doing same algebra as above, we arrive at
\begin{align}\label{Gravitones1}
\widetilde{ \mathcal{M}}(1^{-2}, 2^{-2}, 3^{+2}) =\,& -4i \left(\frac{i}{\sqrt{2}} \right)^2   \times  \biggl\{ (\epsilon^-_g(p_1,q) \cdot p_2) \times ( \epsilon^-_g(p_2,l) \cdot \epsilon^+_g(p_3,r)  ) \nonumber \\
 &~~~+ (\epsilon^-_g(p_2,l) \cdot p_3) \times ( \epsilon^+_g(p_3,r) \cdot \epsilon^-_g(p_1,q)) \biggr\}^2\nonumber \\ 
 &= -4i \left(\frac{i}{\sqrt{2}} \right)^2 \times \left(  -  \frac{ \langle 12 \rangle^3 }{ \sqrt{2}   \langle 23 \rangle \langle 31 \rangle  }      \right)^2 \nonumber\\
 &= i \frac{\langle 12 \rangle^6}{ \langle 23 \rangle^2  \langle 31 \rangle^2 }.
\end{align}
We again applied the Schouten identity (\ref{Schouten}) and momentum conservation (\ref{MCrelation}).
On the other hand, amplitude of two gravitinos one graviton $\widetilde{h} h \widetilde{h}$ with helicities $h_1 = +\frac{3}{2}, h_2 = +2, h_3 = - \frac{3}{2}$ is
\begin{equation}
 \widetilde{ \mathcal{M}}(1^{+\frac{3}{2}}, 2^{+2}, 3^{-\frac{3}{2}}) = V^{ \mu, \alpha \beta, \nu}_{\widetilde{h} h \widetilde{h}}(p_1, p_2, p_3)   \times \widetilde{\epsilon}^+_{\widetilde{h}, \mu}(p_1,q) \epsilon^{++}_{h, \alpha \beta}(p_2,l) \epsilon^-_{\widetilde{h}, \nu}(p_3,r)  
\end{equation}
In the above expression, we are going to use vertex (\ref{Vh}) contracting with tensor fields given by (\ref{h1}),  (\ref{h4}), (\ref{h00}). Using on-shell conditions to further simplify, we obtain the following 
\begin{align}
 \widetilde{ \mathcal{M}}(1^{+\frac{3}{2}}, 2^{+2}, 3^{-\frac{3}{2}}) =\,& i [ \widetilde{\epsilon}^+_{\widetilde{g}}(p_1)  \! \not \epsilon^+_g(p_2,l) \epsilon^-_{\widetilde{g}}(p_3)   \times [ (\epsilon^+_g(p_1,q) \cdot p_2) (\epsilon^-_g(p_3,r) \cdot \epsilon^+_g(p_2,l)) \nonumber\\ 
 & + (\epsilon^+_g(p_2,l)\cdot p_3 )(\epsilon^-_g(p_3,r) \cdot \epsilon^+_g(p_1,q))    ] ],  
\end{align}
where $\widetilde{\epsilon}^+_{\widetilde{g}}(p_1)$ and $\epsilon^-_{\widetilde{g}}(p_3)$ are   gluinos associated to \textcolor{black}{gravitinos} by KLT relations. Next step is to rewrite everything in terms of Weyl helicity spinors, using definitions given by (\ref{e0})  and (\ref{epj0}, \ref{e1e0}),
\begin{align}
\widetilde{ \mathcal{M}}(1^{+\frac{3}{2}}, 2^{+2}, 3^{-\frac{3}{2}}) = \,&i \biggr( [ 1 \vert \left(   \frac{\sqrt{2}}{\langle l2 \rangle } ( \vert 2 ] \langle l \vert + \vert l \rangle [ 2 \vert )  \right) \vert 3 \rangle \nonumber\\
&\times \left( \frac{ [12] \langle 2q \rangle}{ \sqrt{2} \langle q1 \rangle  } \times \frac{\langle 3l \rangle [2r]}{[3r] \langle l2 \rangle}  +  \frac{[23]  \langle 3l \rangle }{ \sqrt{2}  \langle l2 \rangle} \times \frac{\langle 3q \rangle [1r]}{[3r]  \langle q1 \rangle}  \right) \biggr), 
\end{align}
after some algebraic simplification and using $[12] \langle 2q \rangle = [31] \langle 3q \rangle$ and the Schouten Identity for square spinors, we get
\begin{align}
 \widetilde{ \mathcal{M}}(1^{+\frac{3}{2}}, 2^{+2}, 3^{-\frac{3}{2}}) =\,&i \frac{ [12]  \langle l3 \rangle \langle 3q \rangle \langle 3l \rangle}{ \langle l2 \rangle^2 [3r] \langle q1 \rangle}   \times ( - [3r] [12]) \nonumber\\
 =\,&  -i  \frac{[12]^2 \langle l3 \rangle^2 \langle q3 \rangle}{  \langle l2 \rangle^2 \langle q1 \rangle}.
\end{align}
Last step is to apply momentum conservation, 
\begin{equation}
    \langle l2 \rangle [21] + \langle l3 \rangle [31] = 0
\end{equation}
and this implies
\begin{equation}
    \frac{\langle l3 \rangle}{ \langle l2 \rangle} = \frac{[12]}{ [31]},
\end{equation}
and
\begin{equation}
 \langle q1 \rangle [12] + \langle q3 \rangle [32] = 0 ,  
\end{equation}
\begin{equation}
    \frac{\langle q3 \rangle}{ \langle q1 \rangle} = \frac{[12]}{[23]},
\end{equation}
therefore, the final form of the amplitude is
\begin{equation}\label{gravitinos0}
\widetilde{\mathcal{M}}(1^{+\frac{3}{2}}, 2^{+2}, 3^{-\frac{3}{2}}) = i \frac{[12]^5}{ [23] [31]^2}.  
\end{equation}
The next configurations of helicity is $ \widetilde{\mathcal{M}}(1^{-\frac{3}{2}}, 2^{-2}, 3^{+\frac{3}{2}})$, and the corresponding amplitude is
\begin{align}
 \widetilde{ \mathcal{M}}(1^{-\frac{3}{2}}, 2^{-2}, 3^{+\frac{3}{2}}) =\,& i [ \widetilde{\epsilon}^-_{\widetilde{g}}(p_1)  \! \not \epsilon^-_g(p_2,l) \epsilon^+_{\widetilde{g}}(p_3)   \times [ (\epsilon^-_g(p_1,q) \cdot p_2) (\epsilon^-_g(p_2,l) \cdot \epsilon^+_g(p_3,r)) \\ \nonumber
 & + (\epsilon^-_g(p_2,l)\cdot p_3 )(\epsilon^-_g(p_1,q) \cdot \epsilon^+_g(p_3,r))    ] ],  
\end{align}
here we use \eqref{e1} and write everything in terms of Weyl helicity spinors as
\begin{align}
 \widetilde{ \mathcal{M}}(1^{-\frac{3}{2}}, 2^{-2}, 3^{+\frac{3}{2}})  = \,& i \big(  \langle 1 \vert \left( \frac{\sqrt{2}}{ [l2]}  ( \vert 2 \rangle [l \vert + \vert l ] \langle 2 \vert  )  \right) \vert 3 ]\nonumber\\
 &\times      \left(  \frac{ \langle 12 \rangle [2q]}{\sqrt{2} [1q]} \times \frac{\langle 2r \rangle [3l]}{ [2l] \langle r3 \rangle} + \frac{\langle 23 \rangle [3l]}{\sqrt{2} [2l]} \times \frac{\langle 1r \rangle [3q]}{[1q] \langle r3 \rangle}          \right)      \biggr).   
\end{align}
Applying $\langle 12 \rangle [2q] = \langle 31 \rangle [3q]$ and Schouten Identity leads to,
\begin{equation}
\widetilde{ \mathcal{M}}(1^{-\frac{3}{2}}, 2^{-2}, 3^{+\frac{3}{2}}) = i \frac{\langle 12 \rangle^2 [l3]^2 [q3]}{ [l2]^2 [q1]}.   
\end{equation}
According to momentum conservation,
\begin{equation}
    [l2] \langle 21 \rangle + [l3] \langle 31 \rangle = 0
\end{equation}
so
\begin{equation}
    \frac{[l3]}{[l2]} = \frac{\langle 12 \rangle}{ \langle 31 \rangle},
\end{equation}
and $ [q1] \langle 12 \rangle + [q3] \langle 32 \rangle = 0$ implies
\begin{equation}
    \frac{[q3]}{[q1]} = \frac{\langle 12 \rangle}{\langle 23 \rangle }.
\end{equation}
Finally, we arrive at
\begin{equation}\label{gravitinos1}
   \widetilde{ \mathcal{M}}(1^{-\frac{3}{2}}, 2^{-2}, 3^{+\frac{3}{2}}) = i \frac{\langle 12 \rangle^5}{ \langle 23 \rangle \langle 31 \rangle^2}.  
\end{equation}
 We can observe that the process for calculating these three point partial amplitudes does not depend on the choice of a references momentum $q,l$ or $r$.  Suppressing all factor constants, we notice that the results \eqref{Gravitones0}, \eqref{Gravitones1}, \eqref{gravitinos0} and \eqref{gravitinos1} coincide with the results presented in table \eqref{table2} in their spinorial mathematical structure either in terms of $\langle ij \rangle$ or $[ij]$.

\section{ Graviton-Gravitino Compton effect}\label{section5}

Graviton interaction with matter has been studied widely through different points of view over the years. For instance, graviton Compton cross section is calculated using helicity methods in \cite{Holstein:2006bh} and  gravitino physics is studied in different scenarios by \cite{Moroi:1995fs,Diaz-Cruz:2019xjb}. In this section we are interested in computing the total SUGRA Compton amplitude, which is the scattering between  gravitons and  gravitinos. We are going to show that our previous results of the partial amplitudes $\mathcal{\widetilde{M}}$ using  BCFW, in section \eqref{section3}, coincide with the results obtained through the traditional procedure of constructing the amplitudes with the Feynman rules for gravitino interaction from Yang-Mills derived by Bjerrum-Bohr and Engelund \cite{Bjerrum-Bohr:2010eeh}. To obtain the total amplitude for  SUGRA Compton scattering we need the following Feynman Diagrams:
\begin{figure}[H]
    \centering

\tikzset{every picture/.style={line width=0.75pt}} 

\begin{tikzpicture}[x=0.75pt,y=0.75pt,yscale=-1,xscale=1]

\draw    (113,134.82) .. controls (112.89,137.18) and (111.65,138.3) .. (109.3,138.19) .. controls (106.95,138.08) and (105.71,139.2) .. (105.6,141.55) .. controls (105.49,143.9) and (104.25,145.02) .. (101.9,144.91) .. controls (99.55,144.8) and (98.31,145.92) .. (98.2,148.27) .. controls (98.09,150.62) and (96.85,151.75) .. (94.5,151.64) .. controls (92.15,151.53) and (90.91,152.65) .. (90.8,155) .. controls (90.69,157.35) and (89.45,158.47) .. (87.1,158.36) .. controls (84.75,158.25) and (83.51,159.37) .. (83.4,161.72) .. controls (83.29,164.07) and (82.05,165.2) .. (79.7,165.09) .. controls (77.35,164.98) and (76.11,166.1) .. (76,168.45) .. controls (75.89,170.8) and (74.65,171.92) .. (72.3,171.81) -- (68.93,174.87) -- (68.93,174.87)(110.99,132.6) .. controls (110.87,134.95) and (109.63,136.07) .. (107.28,135.96) .. controls (104.93,135.85) and (103.69,136.98) .. (103.58,139.33) .. controls (103.47,141.68) and (102.23,142.8) .. (99.88,142.69) .. controls (97.53,142.58) and (96.29,143.7) .. (96.18,146.05) .. controls (96.07,148.4) and (94.83,149.53) .. (92.48,149.42) .. controls (90.13,149.31) and (88.89,150.43) .. (88.78,152.78) .. controls (88.67,155.13) and (87.43,156.25) .. (85.08,156.14) .. controls (82.73,156.03) and (81.49,157.15) .. (81.38,159.5) .. controls (81.27,161.85) and (80.03,162.98) .. (77.68,162.87) .. controls (75.33,162.76) and (74.09,163.88) .. (73.98,166.23) .. controls (73.87,168.58) and (72.63,169.7) .. (70.28,169.59) -- (66.91,172.65) -- (66.91,172.65) ;
\draw    (111.99,133.71) -- (67.92,93.66) ;
\draw    (209.09,134.36) -- (111.99,133.71) ;
\draw    (256.08,94.13) .. controls (255.95,96.48) and (254.71,97.6) .. (252.36,97.47) .. controls (250.01,97.35) and (248.77,98.47) .. (248.64,100.82) .. controls (248.51,103.17) and (247.27,104.29) .. (244.92,104.16) .. controls (242.57,104.03) and (241.33,105.15) .. (241.21,107.5) .. controls (241.08,109.85) and (239.84,110.97) .. (237.49,110.84) .. controls (235.14,110.72) and (233.9,111.84) .. (233.77,114.19) .. controls (233.64,116.54) and (232.4,117.66) .. (230.05,117.53) .. controls (227.7,117.4) and (226.46,118.52) .. (226.33,120.87) .. controls (226.2,123.22) and (224.96,124.34) .. (222.61,124.21) .. controls (220.26,124.09) and (219.02,125.21) .. (218.9,127.56) .. controls (218.77,129.91) and (217.53,131.03) .. (215.18,130.9) .. controls (212.83,130.77) and (211.59,131.89) .. (211.46,134.24) -- (210.09,135.47) -- (210.09,135.47)(254.07,91.9) .. controls (253.95,94.25) and (252.71,95.37) .. (250.36,95.24) .. controls (248.01,95.12) and (246.77,96.24) .. (246.64,98.59) .. controls (246.51,100.94) and (245.27,102.06) .. (242.92,101.93) .. controls (240.57,101.8) and (239.33,102.92) .. (239.2,105.27) .. controls (239.07,107.62) and (237.83,108.74) .. (235.48,108.61) .. controls (233.13,108.49) and (231.89,109.61) .. (231.76,111.96) .. controls (231.64,114.31) and (230.4,115.43) .. (228.05,115.3) .. controls (225.7,115.17) and (224.46,116.29) .. (224.33,118.64) .. controls (224.2,120.99) and (222.96,122.11) .. (220.61,121.98) .. controls (218.26,121.86) and (217.02,122.98) .. (216.89,125.33) .. controls (216.76,127.68) and (215.52,128.8) .. (213.17,128.67) .. controls (210.82,128.54) and (209.58,129.66) .. (209.45,132.01) -- (208.08,133.24) -- (208.08,133.24) ;
\draw    (256.35,176.35) -- (209.09,134.36) ;
\draw    (401.22,140.59) .. controls (401.08,142.94) and (399.83,144.05) .. (397.48,143.91) .. controls (395.13,143.78) and (393.88,144.89) .. (393.75,147.24) .. controls (393.61,149.59) and (392.36,150.7) .. (390.01,150.56) .. controls (387.66,150.42) and (386.41,151.53) .. (386.28,153.88) .. controls (386.14,156.23) and (384.89,157.34) .. (382.54,157.21) .. controls (380.19,157.07) and (378.94,158.18) .. (378.8,160.53) .. controls (378.67,162.88) and (377.42,163.99) .. (375.07,163.85) .. controls (372.72,163.72) and (371.47,164.83) .. (371.33,167.18) .. controls (371.2,169.53) and (369.95,170.64) .. (367.6,170.5) .. controls (365.25,170.36) and (364,171.47) .. (363.86,173.82) .. controls (363.72,176.17) and (362.47,177.28) .. (360.12,177.14) .. controls (357.77,177.01) and (356.52,178.12) .. (356.39,180.47) -- (354.3,182.33) -- (354.3,182.33)(399.22,138.35) .. controls (399.09,140.7) and (397.84,141.81) .. (395.49,141.67) .. controls (393.14,141.53) and (391.89,142.64) .. (391.75,144.99) .. controls (391.62,147.34) and (390.37,148.45) .. (388.02,148.32) .. controls (385.67,148.18) and (384.42,149.29) .. (384.28,151.64) .. controls (384.15,153.99) and (382.9,155.1) .. (380.55,154.96) .. controls (378.2,154.83) and (376.95,155.94) .. (376.81,158.29) .. controls (376.67,160.64) and (375.42,161.75) .. (373.07,161.61) .. controls (370.72,161.47) and (369.47,162.58) .. (369.34,164.93) .. controls (369.2,167.28) and (367.95,168.39) .. (365.6,168.26) .. controls (363.25,168.12) and (362,169.23) .. (361.87,171.58) .. controls (361.73,173.93) and (360.48,175.04) .. (358.13,174.9) .. controls (355.78,174.77) and (354.53,175.88) .. (354.39,178.23) -- (352.3,180.09) -- (352.3,180.09) ;
\draw    (400.22,139.47) -- (353.3,97.73) ;
\draw    (503.59,140.14) -- (400.22,139.47) ;
\draw    (553.54,98.18) .. controls (553.39,100.53) and (552.13,101.63) .. (549.78,101.48) .. controls (547.43,101.33) and (546.18,102.44) .. (546.03,104.79) .. controls (545.88,107.14) and (544.63,108.24) .. (542.28,108.09) .. controls (539.93,107.94) and (538.67,109.04) .. (538.52,111.39) .. controls (538.37,113.74) and (537.12,114.85) .. (534.77,114.7) .. controls (532.42,114.55) and (531.17,115.65) .. (531.02,118) .. controls (530.87,120.35) and (529.61,121.45) .. (527.26,121.3) .. controls (524.91,121.15) and (523.66,122.26) .. (523.51,124.61) .. controls (523.36,126.96) and (522.11,128.06) .. (519.76,127.91) .. controls (517.41,127.76) and (516.15,128.86) .. (516,131.21) .. controls (515.85,133.56) and (514.6,134.67) .. (512.25,134.52) .. controls (509.9,134.37) and (508.65,135.47) .. (508.5,137.82) .. controls (508.35,140.17) and (507.09,141.27) .. (504.74,141.12) -- (504.58,141.27) -- (504.58,141.27)(551.56,95.93) .. controls (551.41,98.28) and (550.15,99.38) .. (547.8,99.23) .. controls (545.45,99.08) and (544.2,100.19) .. (544.05,102.54) .. controls (543.9,104.89) and (542.65,105.99) .. (540.3,105.84) .. controls (537.95,105.69) and (536.69,106.79) .. (536.54,109.14) .. controls (536.39,111.49) and (535.14,112.59) .. (532.79,112.44) .. controls (530.44,112.29) and (529.18,113.4) .. (529.03,115.75) .. controls (528.88,118.1) and (527.63,119.2) .. (525.28,119.05) .. controls (522.93,118.9) and (521.68,120) .. (521.53,122.35) .. controls (521.38,124.7) and (520.12,125.81) .. (517.77,125.66) .. controls (515.42,125.51) and (514.17,126.61) .. (514.02,128.96) .. controls (513.87,131.31) and (512.62,132.41) .. (510.27,132.26) .. controls (507.92,132.11) and (506.66,133.22) .. (506.51,135.57) .. controls (506.36,137.92) and (505.11,139.02) .. (502.76,138.87) -- (502.59,139.02) -- (502.59,139.02) ;
\draw    (553.91,183.9) -- (503.59,140.14) ;
\draw    (200.68,341.67) .. controls (203.03,341.54) and (204.28,342.65) .. (204.41,345) .. controls (204.54,347.35) and (205.79,348.46) .. (208.14,348.33) .. controls (210.49,348.19) and (211.74,349.3) .. (211.87,351.65) .. controls (212,354) and (213.25,355.11) .. (215.6,354.98) .. controls (217.95,354.85) and (219.2,355.96) .. (219.33,358.31) .. controls (219.46,360.66) and (220.71,361.77) .. (223.06,361.64) .. controls (225.41,361.51) and (226.66,362.62) .. (226.79,364.97) .. controls (226.92,367.32) and (228.17,368.43) .. (230.52,368.3) .. controls (232.87,368.17) and (234.12,369.28) .. (234.25,371.63) .. controls (234.38,373.98) and (235.63,375.09) .. (237.98,374.96) .. controls (240.33,374.83) and (241.58,375.94) .. (241.71,378.29) .. controls (241.84,380.64) and (243.09,381.75) .. (245.44,381.62) -- (247.61,383.55) -- (247.61,383.55)(198.68,343.91) .. controls (201.03,343.77) and (202.28,344.88) .. (202.41,347.23) .. controls (202.54,349.58) and (203.79,350.69) .. (206.14,350.56) .. controls (208.49,350.43) and (209.74,351.54) .. (209.87,353.89) .. controls (210,356.24) and (211.25,357.35) .. (213.6,357.22) .. controls (215.95,357.09) and (217.2,358.2) .. (217.33,360.55) .. controls (217.46,362.9) and (218.71,364.01) .. (221.06,363.88) .. controls (223.41,363.75) and (224.66,364.86) .. (224.79,367.21) .. controls (224.92,369.56) and (226.17,370.67) .. (228.52,370.54) .. controls (230.87,370.41) and (232.12,371.52) .. (232.25,373.87) .. controls (232.38,376.22) and (233.63,377.33) .. (235.98,377.2) .. controls (238.33,377.07) and (239.58,378.18) .. (239.71,380.53) .. controls (239.84,382.88) and (241.09,383.99) .. (243.44,383.86) -- (245.61,385.79) -- (245.61,385.79) ;
\draw    (104.57,342.15) -- (61.39,302.81) ;
\draw    (248.23,301.39) .. controls (248.1,303.74) and (246.86,304.85) .. (244.51,304.72) .. controls (242.16,304.59) and (240.91,305.7) .. (240.78,308.05) .. controls (240.65,310.4) and (239.4,311.51) .. (237.05,311.38) .. controls (234.7,311.25) and (233.45,312.37) .. (233.32,314.72) .. controls (233.19,317.07) and (231.95,318.18) .. (229.6,318.05) .. controls (227.25,317.92) and (226,319.03) .. (225.87,321.38) .. controls (225.74,323.73) and (224.49,324.84) .. (222.14,324.71) .. controls (219.79,324.58) and (218.54,325.7) .. (218.41,328.05) .. controls (218.28,330.4) and (217.04,331.51) .. (214.69,331.38) .. controls (212.34,331.25) and (211.09,332.36) .. (210.96,334.71) .. controls (210.83,337.06) and (209.58,338.17) .. (207.23,338.04) .. controls (204.88,337.91) and (203.63,339.03) .. (203.5,341.38) -- (200.68,343.9) -- (200.68,343.9)(246.23,299.15) .. controls (246.1,301.5) and (244.86,302.61) .. (242.51,302.48) .. controls (240.16,302.35) and (238.91,303.47) .. (238.78,305.82) .. controls (238.65,308.17) and (237.4,309.28) .. (235.05,309.15) .. controls (232.7,309.02) and (231.45,310.13) .. (231.32,312.48) .. controls (231.19,314.83) and (229.95,315.94) .. (227.6,315.81) .. controls (225.25,315.68) and (224,316.8) .. (223.87,319.15) .. controls (223.74,321.5) and (222.49,322.61) .. (220.14,322.48) .. controls (217.79,322.35) and (216.54,323.46) .. (216.41,325.81) .. controls (216.28,328.16) and (215.04,329.27) .. (212.69,329.14) .. controls (210.34,329.01) and (209.09,330.13) .. (208.96,332.48) .. controls (208.83,334.83) and (207.58,335.94) .. (205.23,335.81) .. controls (202.88,335.68) and (201.63,336.79) .. (201.5,339.14) -- (198.68,341.67) -- (198.68,341.67) ;
\draw    (61.39,382.76) -- (104.57,342.15) ;
\draw    (104.58,340.65) .. controls (106.25,339) and (107.92,339.01) .. (109.58,340.68) .. controls (111.23,342.36) and (112.9,342.37) .. (114.58,340.72) .. controls (116.25,339.07) and (117.92,339.08) .. (119.58,340.75) .. controls (121.24,342.42) and (122.91,342.43) .. (124.58,340.78) .. controls (126.26,339.13) and (127.93,339.14) .. (129.58,340.82) .. controls (131.24,342.49) and (132.91,342.5) .. (134.58,340.85) .. controls (136.25,339.2) and (137.92,339.21) .. (139.58,340.88) .. controls (141.23,342.56) and (142.9,342.57) .. (144.58,340.92) .. controls (146.25,339.27) and (147.92,339.28) .. (149.58,340.95) .. controls (151.24,342.62) and (152.91,342.63) .. (154.58,340.98) .. controls (156.26,339.33) and (157.93,339.34) .. (159.58,341.02) .. controls (161.24,342.69) and (162.91,342.7) .. (164.58,341.05) .. controls (166.26,339.4) and (167.93,339.41) .. (169.58,341.09) .. controls (171.24,342.76) and (172.91,342.77) .. (174.58,341.12) .. controls (176.25,339.47) and (177.92,339.48) .. (179.58,341.15) .. controls (181.23,342.83) and (182.9,342.84) .. (184.58,341.19) .. controls (186.25,339.54) and (187.92,339.55) .. (189.58,341.22) .. controls (191.24,342.89) and (192.91,342.9) .. (194.58,341.25) .. controls (196.26,339.6) and (197.93,339.61) .. (199.58,341.29) -- (199.69,341.29) -- (199.69,341.29)(104.56,343.65) .. controls (106.23,342) and (107.9,342.01) .. (109.56,343.68) .. controls (111.21,345.36) and (112.88,345.37) .. (114.56,343.72) .. controls (116.23,342.07) and (117.9,342.08) .. (119.56,343.75) .. controls (121.22,345.42) and (122.89,345.43) .. (124.56,343.78) .. controls (126.24,342.13) and (127.91,342.14) .. (129.56,343.82) .. controls (131.22,345.49) and (132.89,345.5) .. (134.56,343.85) .. controls (136.23,342.2) and (137.9,342.21) .. (139.56,343.88) .. controls (141.21,345.56) and (142.88,345.57) .. (144.56,343.92) .. controls (146.23,342.27) and (147.9,342.28) .. (149.56,343.95) .. controls (151.22,345.62) and (152.89,345.63) .. (154.56,343.98) .. controls (156.24,342.33) and (157.91,342.34) .. (159.56,344.02) .. controls (161.22,345.69) and (162.89,345.7) .. (164.56,344.05) .. controls (166.23,342.4) and (167.9,342.41) .. (169.56,344.08) .. controls (171.21,345.76) and (172.88,345.77) .. (174.56,344.12) .. controls (176.23,342.47) and (177.9,342.48) .. (179.56,344.15) .. controls (181.21,345.83) and (182.88,345.84) .. (184.56,344.19) .. controls (186.23,342.54) and (187.9,342.55) .. (189.56,344.22) .. controls (191.22,345.89) and (192.89,345.9) .. (194.56,344.25) .. controls (196.24,342.6) and (197.91,342.61) .. (199.56,344.29) -- (199.67,344.29) -- (199.67,344.29) ;
\draw    (450.26,346.72) .. controls (452.54,346.14) and (453.97,346.99) .. (454.56,349.27) .. controls (455.15,351.56) and (456.58,352.41) .. (458.87,351.82) .. controls (461.15,351.23) and (462.58,352.08) .. (463.17,354.36) .. controls (463.75,356.65) and (465.18,357.5) .. (467.47,356.91) .. controls (469.75,356.32) and (471.19,357.17) .. (471.78,359.45) .. controls (472.36,361.74) and (473.79,362.59) .. (476.08,362) .. controls (478.36,361.41) and (479.79,362.26) .. (480.38,364.54) .. controls (480.97,366.83) and (482.4,367.68) .. (484.69,367.09) .. controls (486.97,366.5) and (488.4,367.35) .. (488.99,369.63) .. controls (489.58,371.92) and (491.01,372.77) .. (493.3,372.18) .. controls (495.58,371.59) and (497.01,372.44) .. (497.6,374.72) .. controls (498.18,377.01) and (499.61,377.86) .. (501.9,377.27) .. controls (504.18,376.68) and (505.62,377.53) .. (506.21,379.81) .. controls (506.79,382.1) and (508.22,382.95) .. (510.51,382.36) .. controls (512.79,381.77) and (514.22,382.62) .. (514.81,384.9) -- (515.98,385.59) -- (515.98,385.59)(448.73,349.31) .. controls (451.02,348.72) and (452.45,349.57) .. (453.04,351.85) .. controls (453.62,354.14) and (455.05,354.99) .. (457.34,354.4) .. controls (459.62,353.81) and (461.05,354.66) .. (461.64,356.94) .. controls (462.23,359.23) and (463.66,360.08) .. (465.95,359.49) .. controls (468.23,358.9) and (469.66,359.75) .. (470.25,362.03) .. controls (470.83,364.32) and (472.26,365.17) .. (474.55,364.58) .. controls (476.83,363.99) and (478.27,364.84) .. (478.86,367.12) .. controls (479.44,369.41) and (480.87,370.26) .. (483.16,369.67) .. controls (485.44,369.08) and (486.87,369.93) .. (487.46,372.21) .. controls (488.05,374.5) and (489.48,375.35) .. (491.77,374.76) .. controls (494.05,374.17) and (495.48,375.02) .. (496.07,377.3) .. controls (496.66,379.59) and (498.09,380.44) .. (500.38,379.85) .. controls (502.66,379.26) and (504.09,380.11) .. (504.68,382.39) .. controls (505.26,384.68) and (506.69,385.53) .. (508.98,384.94) .. controls (511.26,384.35) and (512.7,385.2) .. (513.29,387.48) -- (514.45,388.17) -- (514.45,388.17) ;
\draw    (449.49,348.02) -- (389.03,311.5) ;
\draw    (516.86,309.85) .. controls (516.27,312.14) and (514.84,312.99) .. (512.56,312.4) .. controls (510.27,311.81) and (508.84,312.66) .. (508.25,314.95) .. controls (507.67,317.24) and (506.24,318.09) .. (503.95,317.5) .. controls (501.67,316.91) and (500.24,317.76) .. (499.65,320.04) .. controls (499.07,322.33) and (497.64,323.18) .. (495.35,322.59) .. controls (493.06,322) and (491.63,322.85) .. (491.05,325.14) .. controls (490.47,327.43) and (489.04,328.28) .. (486.75,327.69) .. controls (484.46,327.1) and (483.03,327.95) .. (482.44,330.24) .. controls (481.86,332.53) and (480.43,333.38) .. (478.14,332.79) .. controls (475.85,332.2) and (474.42,333.05) .. (473.84,335.34) .. controls (473.25,337.62) and (471.82,338.47) .. (469.54,337.88) .. controls (467.25,337.29) and (465.82,338.14) .. (465.24,340.43) .. controls (464.65,342.72) and (463.22,343.57) .. (460.93,342.98) .. controls (458.64,342.39) and (457.21,343.24) .. (456.63,345.53) .. controls (456.05,347.82) and (454.62,348.67) .. (452.33,348.08) -- (450.26,349.31) -- (450.26,349.31)(515.33,307.27) .. controls (514.74,309.55) and (513.31,310.4) .. (511.03,309.82) .. controls (508.74,309.23) and (507.31,310.08) .. (506.72,312.37) .. controls (506.14,314.66) and (504.71,315.51) .. (502.42,314.92) .. controls (500.14,314.33) and (498.71,315.18) .. (498.12,317.46) .. controls (497.54,319.75) and (496.11,320.6) .. (493.82,320.01) .. controls (491.53,319.42) and (490.1,320.27) .. (489.52,322.56) .. controls (488.94,324.85) and (487.51,325.7) .. (485.22,325.11) .. controls (482.93,324.52) and (481.5,325.37) .. (480.91,327.66) .. controls (480.33,329.95) and (478.9,330.8) .. (476.61,330.21) .. controls (474.33,329.62) and (472.9,330.47) .. (472.31,332.75) .. controls (471.73,335.04) and (470.3,335.89) .. (468.01,335.3) .. controls (465.72,334.71) and (464.29,335.56) .. (463.71,337.85) .. controls (463.13,340.14) and (461.7,340.99) .. (459.41,340.4) .. controls (457.12,339.81) and (455.69,340.66) .. (455.1,342.95) .. controls (454.52,345.24) and (453.09,346.09) .. (450.8,345.5) -- (448.73,346.73) -- (448.73,346.73) ;
\draw    (389.03,385.71) -- (449.49,348.02) ;

\draw (46.62,68.37) node [anchor=north west][inner sep=0.75pt]    {$1_{\tilde{h}}^{-}$};
\draw (46.25,174.09) node [anchor=north west][inner sep=0.75pt]    {$4_{h}^{+}$};
\draw (257.66,71.84) node [anchor=north west][inner sep=0.75pt]    {$2_{h}^{-}$};
\draw (258.35,179.75) node [anchor=north west][inner sep=0.75pt]    {$3_{\tilde{h}}^{+}$};
\draw (121.99,195.3) node [anchor=north west][inner sep=0.75pt]   [align=left] {Diagram A};
\draw (331.46,71.03) node [anchor=north west][inner sep=0.75pt]    {$1_{\tilde{h}}^{-}$};
\draw (554.71,75.54) node [anchor=north west][inner sep=0.75pt]    {$4_{h}^{+}$};
\draw (333.14,184.52) node [anchor=north west][inner sep=0.75pt]    {$2_{h}^{-}$};
\draw (555.91,187.3) node [anchor=north west][inner sep=0.75pt]    {$3_{\tilde{h}}^{+}$};
\draw (413.02,204.01) node [anchor=north west][inner sep=0.75pt]   [align=left] {Diagram B};
\draw (147.9,139.86) node [anchor=north west][inner sep=0.75pt]    {$s_{14}$};
\draw (439.86,143.6) node [anchor=north west][inner sep=0.75pt]    {$s_{12}$};
\draw (41.51,383.32) node [anchor=north west][inner sep=0.75pt]    {$1_{\tilde{h}}^{-}$};
\draw (248.61,388.07) node [anchor=north west][inner sep=0.75pt]    {$4_{h}^{+}$};
\draw (250.12,282.14) node [anchor=north west][inner sep=0.75pt]    {$2_{h}^{-}$};
\draw (39.88,278.32) node [anchor=north west][inner sep=0.75pt]    {$3_{\tilde{h}}^{+}$};
\draw (113.25,402.5) node [anchor=north west][inner sep=0.75pt]   [align=left] {Diagram C};
\draw (139.37,353.44) node [anchor=north west][inner sep=0.75pt]    {$s_{13}$};
\draw (369.52,386.76) node [anchor=north west][inner sep=0.75pt]    {$1_{\tilde{h}}^{-}$};
\draw (517.22,390.28) node [anchor=north west][inner sep=0.75pt]    {$4_{h}^{+}$};
\draw (519.36,284.69) node [anchor=north west][inner sep=0.75pt]    {$2_{h}^{-}$};
\draw (366.64,284.25) node [anchor=north west][inner sep=0.75pt]    {$3_{\tilde{h}}^{+}$};
\draw (421.26,402.23) node [anchor=north west][inner sep=0.75pt]   [align=left] {Diagram D};
\end{tikzpicture}
 \caption{Feynman Diagrams}
    \label{Feynman Diagrams}
\end{figure}
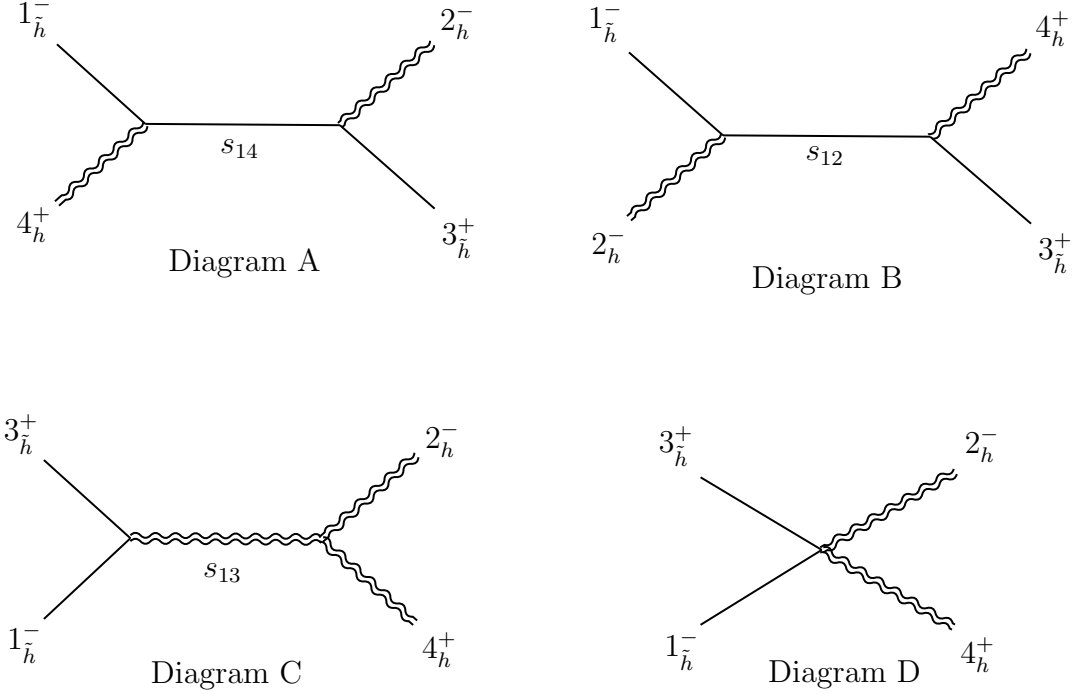
where we have fixed the standard helicity configuration for each diagram, $\mathcal{\Tilde{M}} (1^-_{\Tilde{h}}, 2^-_h, 3^+_{\Tilde{h}}, 4^+_h)$. Also we need two propagators-- one for the gravitino and one for the graviton. The propagator for the gravitino is chosen as:
\begin{equation}\label{propagator gravitino}
    P_{\widetilde{h} \widetilde{h}, \mu \nu}(p) = \frac{-i \! \not p \eta_{\mu \nu}}{ p^2 + i\epsilon},
\end{equation}
and for gravitons:
\begin{equation}\label{propagator for graviton}
    P_{hh, \mu \alpha, \nu \beta }(p) = \frac{i \eta_{\mu \nu}  \eta_{\alpha \beta}}{p^2 + i \epsilon}.
\end{equation}
This is same as in \cite{Bjerrum-Bohr:2010eeh, Das:1976ct}.
\subsection{Diagram D and the four-point vertex}
We begin to compute the amplitude of diagram D with four point vertex.  According to Bjerrum-Bohr and Engelund \cite{Bjerrum-Bohr:2010eeh}, the four-point vertex for two gravitinos and two gravitons is 
\begin{align}
  V_{\widetilde{h} \widetilde{h} h h}^{\mu, \nu, \alpha \beta, \kappa \lambda} (p_1,p_3, p_2, p_4) =\,& \frac{i}{8} \biggl\{   3 (\eta^{\mu \beta}  \eta^{\nu \lambda} - \eta^{\mu \lambda} \eta^{\nu \beta}) [ (\! \not p_2  - \! \not p_4) \eta^{\alpha \kappa} + \gamma^{\alpha} ( -2p_2 - p_4)^{\kappa}  \nonumber\\ 
 &\, + \gamma^{\kappa}(2 p_4 + p_2)^{\alpha}]- (\gamma^{\kappa}(\! \not p_4  + \! \not p_1) \gamma^{\alpha} + \gamma^{\alpha}(\! \not p_1 + \! \not p_2) \gamma^{\kappa}) \nonumber\\
 &\, ( 2 \eta^{\mu \nu} \eta^{\beta \lambda} - \eta^{\mu \beta} \eta^{\nu \lambda} - \eta^{\mu \lambda} \eta^{\nu \beta} ) \biggr\}. 
\end{align}
Contracting this four vertex with polarization tensors (\ref{h2}, \ref{h3}, \ref{h11}, \ref{h00}), leads to the following partial amplitude
\begin{align}\label{M4D}
\widetilde{\mathcal{M}}_D (1^{-\frac{3}{2}}, 2^{-2}, 3^{+ \frac{3}{2}}, 4^{-2}) =\,& V_{\widetilde{h} \widetilde{h} h h}^{\mu, \nu, \alpha \beta, \kappa \lambda} (p_1,p_3, p_2, p_4) \times \widetilde{\epsilon}^-_{\widetilde{h}, \mu}(p_1,q)  \epsilon^+_{\widetilde{h},\nu}(p_3,r) \epsilon^{- -}_{h, \alpha \beta}(p_2,l) \epsilon^{++}_{h, \kappa \lambda}(p_4,j) \nonumber\\ 
 =\,& \frac{i}{8} \biggl\{  -3 ( \epsilon^-_g(p_1, q) \cdot \epsilon_g^+(p_4, j)) ( \epsilon^+_g(p_3, r) \cdot \epsilon^-_g(p_2, l)) \times \widetilde{\epsilon}^-_{\widetilde{g}}(p_1, q) \epsilon^+_{\widetilde{g}}(p_3, r) \nonumber \\ 
\,&\times   [   (\! \not p_2 - \! \not p_4)(\epsilon^-_g(p_2, l) \cdot \epsilon^+_g(p_4, j)) + \! \not \epsilon^-_g(p_2, l)\times( -2 \epsilon^+_g(p_4, j) \cdot p_2)\nonumber\\
\,\,&  + \! \not \epsilon^+_g(p_4, j)\times (2 \epsilon^-_g(p_2, l) \cdot p_4) ] \nonumber\\ 
\,&- [  \! \not \epsilon^+_g(p_4, j) (\! \not p_4 + \! \not p_1) \! \not \epsilon^-_g(p_2, l) + \! \not \epsilon_g^-(p_2, l) (\! \not p_1 + \! \not p_2) \! \not \epsilon^+_g(p_4, j)] \nonumber\\ 
\,&\times [2 (\epsilon^-_g(p_1, q) \cdot \epsilon^+_g(p_3, r)) - (\epsilon^-_g(p_1, q) \cdot \epsilon^+_g(p_4, j)) (\epsilon^+_g(p_3, r) \cdot \epsilon^-_g(p_2, l))]\nonumber \\ 
& \times  \widetilde{\epsilon}^-_{\widetilde{g}}(p_1, q)  \epsilon^+_{\widetilde{g}}(p_3, r)  \biggr\}.
\end{align}
where we are rewriting polarization tensors of gravitons and gravitinos in terms of polarization vectors of gluons and gluinos due to KLT gauge theory-gravity relations. Here $\widetilde{\epsilon}^-_{\widetilde{g}}(p_1) = \langle 1 \vert$ and $\epsilon^+_{\widetilde{g}}(p_3) = \vert 3 ]$. We choose references momentum of particles with momentum $p_1$ and $p_2$ as $q=l= 4$, particles with momentum $p_3$ and $p_4$ as $r=j = 1$ \footnote{We shall use this choice of reference momentum for all Feynman diagrams}, using this information on equations (\ref{e1} - \ref{e1e0}), all vanish except the next polarization vector contractions:
\begin{equation}
    \epsilon^-_g(p_2, l) \cdot \epsilon_g^+(p_3,r) =\frac{ \langle 21 \rangle [34]}{[24] \langle 13 \rangle},
\end{equation}
\begin{equation}
    \epsilon^+_g(p_2,l) \cdot p_1 = \frac{[42] \langle 21 \rangle}{ \sqrt{2} \langle 14 \rangle}.
\end{equation}
The terms in the amplitude (\ref{M4D}) which become zero, are the following:
\begin{align}
\epsilon^-_g(p_1,q) \cdot \epsilon_g^+(p_4,j) = \epsilon^-_g(p_2,l) \cdot \epsilon^+_g(p_4,j) = \epsilon^-_g (p_1,q) \cdot \epsilon^+_g(p_3,r) = \epsilon^-_g(p_2,l) \cdot p_4 = 0,    
\end{align}
 therefore
\begin{equation}
 \widetilde{\mathcal{M}}_D (1^{-\frac{3}{2}}, 2^{-2}, 3^{+ \frac{3}{2}}, 4^{-2}) = 0.  
\end{equation}
\subsection{Feynman Diagram A}
The partial amplitude for the $s_{14}$ channel according to Feynman rules (where we do the contraction between polarization tensors, three point vertices and propagators) is given by
\begin{align}
 \mathcal{\widetilde{M}}_{A}(1^{-\frac{3}{2}}, 2^{-2}, 3^{+ \frac{3}{2}}, 4^{-2})  =\,& \widetilde{\epsilon}^-_{\widetilde{h}, \kappa}(p_1,q) V^{\kappa, \alpha \beta, \mu}_{\widetilde{h} h \widetilde{h}}(p_1, p_4, p) \epsilon^{++}_{h, \alpha \beta}(p_4,j) \times P_{\widetilde{h} \widetilde{h}, \mu \nu}(p) \nonumber\\ 
 & \times \epsilon^+_{\widetilde{h}, \tau}(p_3, r) V^{\tau, \rho \sigma, \nu}_{\widetilde{h} h \widetilde{h}}(p_3,p_2,p) \epsilon^{--}_{h, \rho \sigma}(p_2,l) \nonumber\\ 
  =\,&  \widetilde{\epsilon}_{\widetilde{g}}^-(p_1) \biggl\{  \frac{i}{2}  \! \not \epsilon_g^+(p_4,j) [ (2p_1 + p)_{\nu} (\epsilon_g^-(p_1, q) \cdot \epsilon^+_g(p_4, j)) \nonumber\\ 
 &+ 2 ( \epsilon_g^-(p_1, q) \cdot p_4) \epsilon^+_{g, \nu}(p_4, j)   + 2(\epsilon^+_g(p_4, j) \cdot p) \epsilon^-_{g, \nu}(p_1, q)   ]  \biggr\} \nonumber \\ 
&\times \left(  
\frac{-i \! \not p}{s_{14}}  \right)\times \epsilon^+_{\widetilde{g}}(p_3) \biggl\{   \frac{i}{2}   \! \not \epsilon_g^-(p_2, l)  [  (2p_3 + p)^{\nu} (\epsilon^-_g(p_2, l) \cdot \epsilon_g^+(p_3, r))\nonumber \\ 
&+ 2(\epsilon_g^+(p_3, r) \cdot p_2) \epsilon_g^{-, \nu}(p_2, l) + 2 (\epsilon_g^-(p_2, l) \cdot p) \epsilon^{+, \nu}_g(p_3, r)]   \biggr\}.
\end{align}
In the first equality everything is written in gravitino-graviton terms where the gravitino propagator is given by \eqref{propagator gravitino}, the second equality is written in gluon-gluino terms due to KLT gauge theory-gravity relations. Now, due to the presence of these on-shell conditions $ \epsilon^-_g(p_1, q) \cdot \epsilon_g^+(p_4, j) = \epsilon_g^-(p_1, q) \cdot p_4 = \epsilon_g^+(p_4, j) \cdot p =0$, with  $p = p_1 + p_4$, the amplitude vanish
\begin{equation}
 \mathcal{\widetilde{M}}_{4A}(1^{-\frac{3}{2}}, 2^{-2}, 3^{+ \frac{3}{2}}, 4^{-2}) = 0.   
\end{equation}
This result coincide with that obtained in (\ref{MBCFWA}). Furthermore, according to equations (\ref{KLTgluons}) and (\ref{KLTgluinos}), KLT relations tell us that for two gravitinos and two gravitons, we have
\begin{equation}
   \widetilde{\mathcal{M}}_{A s_{14}} (1_{\widetilde{h}}^-, 2_h^-, 3^+_{\widetilde{h}}, 4^+_h) = -i s_{14}  \mathcal{A}_4 (1^-_{\widetilde{g}}, 4^+_g , 2^-_g, 3^+_{\widetilde{g}} ) \times \mathcal{A}_4 (1^-_g, 2^-_g, 3^+_g, 4^+_g) = 0.
\end{equation}
This channel does not contribute at all to the process.
\subsection{Feynman Diagram B}
The partial amplitude for the $s_{12}$ channel using Feynman rules is
 \begin{align}\label{MabF}
     \mathcal{\widetilde{M}}_{B}(1^{-\frac{3}{2}}, 2^{-2}, 3^{+ \frac{3}{2}}, 4^{-2}) =\,& \widetilde{\epsilon}^-_{\widetilde{h}, \kappa} (p_1, q) V^{\kappa, \alpha \beta, \mu}_{\widetilde{h} h \widetilde{h}}(p_1, p_2, p) \epsilon^{--}_{h, \alpha \beta}(p_2,l) \times P_{\widetilde{h} \widetilde{h}, \mu \nu}(p) \nonumber\\  
     \,&\times \epsilon^+_{\widetilde{h}, \tau}(p_3,r) V^{\tau, \rho \sigma, \nu}_{\widetilde{h} h \widetilde{h}}(p_3, p_4,p) \epsilon^{++}_{h, \rho \sigma}(p_4,j)\nonumber \\ 
     =\,&  \widetilde{\epsilon}_{\widetilde{g}}^-(p_1) \! \not \epsilon_g^-(p_2, l) \times \!\not p  \times \!\not \epsilon_g^+(p_4, j) \epsilon^+_{\widetilde{g}}(p_3) \nonumber\\ 
     \,&\times \left( \frac{i}{s_{12}} ( \epsilon_g^-(p_1, q) \cdot p_2) ( \epsilon^+_g (p_4, j) \cdot p_3)( \epsilon^-_g(p_2, l) \cdot \epsilon_g^+(p_3, r)) \right).
 \end{align}
Again, in the first equality everything is in gravitino-graviton terms, the second equality is in gluon-gluino  terms where $ s_{12} = \langle 34 \rangle [43] $,  $\widetilde{\epsilon}^-_{\widetilde{g}}(p_1) = \langle 1 \vert$, $\epsilon^+_{\widetilde{g}}(p_3) = \vert 3 ]$ and the propagator we used is \eqref{propagator gravitino}. According to appendix expressions (\ref{e1} - \ref{e1e0}), We have the following results:
\begin{align}
    \!\not \epsilon_g^-(p_2, l) = \,&\frac{\sqrt{2}}{[42]} ( \vert 2 \rangle [4\vert + \vert 4] \langle 2 \vert),\\
    \!\not \epsilon_g^+(p_4, j) =\,& \frac{\sqrt{2}}{\langle 14 \rangle} ( \vert 4] \langle 1 \vert + \vert 1 \rangle [4 \vert ),\\
    \!\not p \, =\,  \!\not p_1 + \!\not p_2 =\,&\vert 1 \rangle [1 \vert + \vert 1] \langle 1 \vert + \vert 2 \rangle [2 \vert + \vert 2] \langle 2 \vert,
\end{align}
\begin{equation}
 ( \epsilon_g^-(p_1, q) \cdot p_2) =    \frac{\langle 12 \rangle [24]}{\sqrt{2} [14]}, \quad  ( \epsilon^+_g (p_4, j) \cdot p_3) = \frac{[43] \langle 31 \rangle}{ \sqrt{2} \langle 14 \rangle}, \quad  ( \epsilon^-_g(p_2, l) \cdot \epsilon_g^+(p_3, r)) = \frac{\langle 21 \rangle [34]}{[24] \langle 13 \rangle}.
\end{equation}
Recall that our reference momenta are $q = l = 4$ and $r = j = 1$. Then substituting all these  in (\ref{MabF}), the final answer is
\begin{align}\label{FDB}
    \mathcal{\widetilde{M}}_{B}(1^{-\frac{3}{2}}, 2^{-2}, 3^{+ \frac{3}{2}}, 4^{-2}) =\,& i \frac{\langle 12 \rangle^2 [43] \langle 12 \rangle^2 [43]^2}{ \langle 34 \rangle [43] \langle 14 \rangle^2 [14]} \\ \nonumber
    =\,& i \frac{ \langle 12 \rangle^6 [34]}{\langle 12 \rangle \langle 23 \rangle \langle 34 \rangle \langle 41 \rangle^2}.
\end{align}
Where
\begin{equation*}
    \frac{[34]}{[14]} = - \frac{\langle 21 \rangle}{ \langle 23 \rangle}.
\end{equation*}
We can see that the result of (\ref{FDB}) coincides with that obtained in (\ref{MBCFWB}) by using BCFW relations. Furthermore, using KLT relations for two gravitinos and two gravitons, for this diagram B, we get
\begin{align}
\mathcal{\widetilde{M}}_{B s_{13}}(1_{\widetilde{h}}^-, 2_h^-, 3^+_{\widetilde{h}}, 4^+_h) &= - i s_{13} \mathcal{A}_4 (1^-_{\widetilde{g}}, 3^+_{\widetilde{g}}, 2^-_g, 4^+_g ) \times \mathcal{A}_4 (1^-_g, 2^-_g, 3^+_g, 4^+_g) \\ \nonumber
&= -i s_{13} \frac{\langle 21\rangle^3}{ \langle 14 \rangle \langle 42 \rangle \langle 31 \rangle} \times \frac{\langle 12 \rangle^4}{ \langle 12 \rangle \langle 23 \rangle \langle 34 \rangle \langle 41 \rangle} \\ \nonumber
&= -i \frac{\langle 12 \rangle^6 [34]}{\langle 12 \rangle  \langle 23 \rangle \langle 34 \rangle \langle 41 \rangle^2},
\end{align}
where $s_{13}= \langle 13 \rangle [31]$ and $\langle 21 \rangle [13] = \langle 42 \rangle [43]$.

\subsection{Feynman Diagram C}

Following the same procedure as above, the partial amplitude for $s_{13}$ channel is
\begin{align}
 \mathcal{\widetilde{M}}_{C}(1^{-\frac{3}{2}}, 2^{-2}, 3^{+ \frac{3}{2}}, 4^{-2})    =\,& \widetilde{\epsilon}_{\widetilde{h}, \kappa}^-(p_1,q) V^{\kappa, \mu \theta, \tau}_{\widetilde{h} h \widetilde{h}}(p_1,p,p_3) \epsilon^+_{\widetilde{h}, \tau}(p_3,r) \times P_{hh, \mu \theta, \nu \gamma}(p) \nonumber\\ 
\, & \times \epsilon^{--}_{h, \rho \sigma}(p_2,l) V^{\rho \sigma, \nu \gamma, \alpha \beta}_{hhh}(p_2,p,p_4) \epsilon^{++}_{h, \alpha \beta}(p_4,j) \nonumber\\ 
  =\,& \frac{-i}{4 s_{13}} \widetilde{\epsilon}_{\widetilde{g}}^-(p_1)\nonumber\\
  \,& \biggl\{ 8 ( \epsilon^-_g(p_1, q) \cdot p_3) ( \epsilon^-_g(p_2, l) \cdot \epsilon^+_g(p_3, r) )  (\epsilon^+_g(p_4, j) \cdot p_2)^2 \!\not \epsilon^-_g(p_2, l) \biggr\} \nonumber\\
  \,& \epsilon^+_{\widetilde{g}}(p_3),
\end{align}
Again, due to KLT gauge theory-gravity relations we rewrite the amplitude in gluon-gluino terms, where $s_{13} = \langle 24 \rangle [42]$ and using the following results:
\begin{equation}
 ( \epsilon^-_g(p_1) \cdot p_3)= \frac{\langle 13 \rangle [34]}{\sqrt{2} [14]}, \quad ( \epsilon^-_g(p_2) \cdot \epsilon^+_g(p_3) )= \frac{\langle 21 \rangle [34]}{[24] \langle 13 \rangle}, \quad    (\epsilon^+_g(p_4) \cdot p_2) = \frac{[42] \langle 21 \rangle }{\sqrt{2} \langle 14 \rangle},
\end{equation}
\begin{equation}
 \widetilde{\epsilon}_{\widetilde{g}}^-(p_1)   \!\not \epsilon^-_g(p_2) \epsilon^+_{\widetilde{g}}(p_3) = \frac{\sqrt{2}  \langle 12 \rangle [43]}{[42]},
\end{equation}
we get 
\begin{align}\label{McFR}
  \mathcal{\widetilde{M}}_{C}(1^{-\frac{3}{2}}, 2^{-2}, 3^{+ \frac{3}{2}}, 4^{-2}) &= - i \frac{\langle 12 \rangle^4 [34]^3 [42]^2}{ \langle 24 \rangle [42] [14] [42] \langle 14 \rangle^2 [42]} \\ \nonumber
  &=i \frac{\langle 12 \rangle^6 [34]}{ \langle 13 \rangle \langle 32 \rangle \langle 24 \rangle \langle 41\rangle^2}.   
\end{align}
Furthermore, applying KLT in diagram C, we get
\begin{align}
\mathcal{\widetilde{M}}_{C s_{12}}(1_{\widetilde{h}}^-, 2_h^-, 3^+_{\widetilde{h}}, 4^+_h) &= - i s_{12} \mathcal{A}_4 (1^-_{\widetilde{g}},2^-_g, 3^+_{\widetilde{g}}, 4^+_g ) \times \mathcal{A}_4 (1^-_g, 3^+_g, 2^-_g, 4^+_g)\nonumber \\ 
&= -i s_{12} \frac{\langle 12 \rangle^3}{ \langle 12 \rangle \langle 34 \rangle \langle 41 \rangle} \times \frac{\langle 21 \rangle^4}{ \langle 14 \rangle \langle 42 \rangle \langle 23 \rangle \langle 31 \rangle} \nonumber\\ 
&= i \frac{\langle 12 \rangle^6 [34]}{\langle 13 \rangle  \langle 32 \rangle \langle 24 \rangle \langle 41 \rangle^2},
\end{align}
with $s_{12} = \langle 34 \rangle [43]$. Notice that, the expression (\ref{McFR}) coincides with that obtained using BCFW procedure (\ref{MBCFWC}). Finally, the total amplitude for the SUGRA Compton scattering becomes,
\begin{align}
\mathcal{M}_{Total} &= \textcolor{black}{ \widetilde{g}_{123} g_{123}} \widetilde{\mathcal{M}}_B + \textcolor{black}{g_{123} \widetilde{G}_{123}} \widetilde{\mathcal{M}}_C \nonumber\\ 
& = \textcolor{black}{ \widetilde{g}_{123} g_{123}} \left(  -i \frac{\langle 12 \rangle^6 [34]}{ \langle 12 \rangle \langle 23 \rangle \langle 34 \rangle  \langle 41 \rangle^2} \right) + \textcolor{black}{g_{123} \widetilde{G}_{123}}  \left(i \frac{\langle 12 \rangle^6 [34]}{ \langle 13 \rangle  \langle 32 \rangle \langle 24 \rangle \langle 41 \rangle^2}\right) \nonumber \\
&= i \langle 12 \rangle [34] \left(  \textcolor{black}{g_{123} \widetilde{G}_{123}}   \frac{\langle 12 \rangle^5 }{ \langle 13 \rangle  \langle 32 \rangle \langle 24 \rangle \langle 41 \rangle^2}   -   \textcolor{black}{ \widetilde{g}_{123} g_{123}}  \frac{\langle 12 \rangle^5}{ \langle 12 \rangle \langle 23 \rangle \langle 34 \rangle  \langle 41 \rangle^2}    \right)
\end{align}

\section{Conclusions}

In this paper, we have shown that our results for three and four point amplitudes  for gravitino and graviton interactions in $\mathcal{N} =1$ SUGRA, coincide with both on-shell methods and conventional analysis of Feynman rules. For three particle amplitudes, at first, we applied  the master formula, which is obtained in the literature from little group properties \cite{Benincasa:2007xk, Cheung:2017pzi}. Using this formula, we have two helicity configurations corresponding to two different three point amplitudes. One is for gravitino-gravitino-graviton interaction and the other one for graviton-graviton-graviton interaction. Suppressing all the factors of the coupling constants,  we show that these interactions possess the same mathematical structure as those  obtained by contracting three point KLT vertices for gravitino-gravitino-graviton interaction and graviton-graviton-graviton interaction,  derived by Bjerrum-Bohr and Engelund \cite{Bjerrum-Bohr:2010eeh}, with their corresponding polarization tensors which can be written in terms of massless helicity spinors.  To get the four point amplitudes for SUGRA Compton scattering from the three point ones we make use of BFCW techniques which is part of the modern on shell methods. Interestingly, the results match if we use KLT Feynman rules \cite{Bjerrum-Bohr:2010eeh} to calculate the amplitudes for every Feynman diagram then rewrite everything in terms of helicity spinors. Finally, we emphasize  that it is easier to calculate amplitudes even for a complicated theory as supergravity by means of  modern on shell methods than conventional analysis of Feynman rules.

\bigskip

\noindent\textbf{Acknowledgments}:
JLDC would like to thank the support of CONACYT and SNI (Mexico). DC would like to thank VIEP-BUAP for the postdoctoral fellowship. The work of JRP is supported by the CONAHCYT scholarship  \textit{Becas Nacional  (Tradicional) 2022 - 1}.

\bigskip

\begin{appendix}

\section{Helicity spinors formalism}
Over the years, it has been proven to be useful to describe massless particles in scattering processes through the well known helicity spinors that are defined as real or complex doublets, transforming under $(\frac{1}{2},0 )$ or $(0, \frac{1}{2})$ representations of the Lorentz group \cite{Kleiss:1985yh, Hagiwara:1985yu, Badger:2023eqz}. To represent the momenta as bispinors, we use sigma Pauli matrices \cite{Schwartz:2014sze} :
	\begin{align}\label{psigma}
	    p^{\dot{a} a}& =(   \bar{ \sigma}^{\mu})^{\dot{a} a} p_{\mu} =	\left(
		\begin{array}{cc}
			p^0 - p^3 &  -p^1 + i p^2  \\
			-p^1 - i p^2 &  p^0 + p^3   
		\end{array} \right) \\ \nonumber
  &=  \lambda_a   \widetilde{ \lambda}_{\Dot{a}} \equiv \lvert p \rangle [ p \lvert
	\end{align}
Dirac spinors can be either left or right handed, and in the Weyl basis we can write them as:
\begin{equation}
		v_+(p) = u_-(p)=	\left(
		\begin{array}{c}
			\lvert p ]_a \\
			0
		\end{array} \right) \equiv \lvert p ] , \quad
		v_-(p)= u_+(p) =	\left(
		\begin{array}{c}
			0 \\
			\lvert p \rangle^{\dot{a}} 
		\end{array} \right) \equiv \lvert p \rangle, 
	\end{equation} 
	\begin{equation}
		\bar{u}_-(p) = \bar{v}_+(p) =	( 0 , \langle p \lvert_{\dot{a}} ) \equiv \langle p \lvert ,  \quad  \bar{u}_+(p)  
= \bar{v}_-(p) = ( [ p \lvert^{a} , 0) \equiv [ p \lvert.
	\end{equation}
This is possible in the high energy limit where the angle and square brackets represent Weyl spinors, i.e, two component spinors which are not independent of each other. At the same time, they can be considered as the building blocks of the the massless particles \cite{Diaz-Cruz:2015oie}. The familiar Mandelstam invariants in terms of angle and square brackets are
\begin{equation}
    s_{ij} = (p_i + p_j)^2 =  2(p_i \cdot p_j) = \langle ij \rangle [ji].
\end{equation}
Some important algebraic properties of helicity spinors are given by the Schouten identity and momentum conservation,
\begin{equation}\label{Schouten}
\langle 12 \rangle \langle 34 \rangle + \langle 13 \rangle \langle 42 \rangle + \langle 14 \rangle \langle 23 \rangle = 0,
\end{equation}
\begin{equation}\label{MCrelation}
 \sum^n_{i=1} \langle k i \rangle [i j] = 0.
\end{equation}
from which we benefit in reducing the rigorous numerical calculations. Meanwhile, boson polarization vectors can also be written in terms of Weyl spinors in the same way as we did for 4-momentum vector  (\ref{psigma}). We are interested in the contractions between polarization vectors and momenta, in terms of Weyl spinors, which we write below \cite{Schwartz:2014sze,Xu:1986xb}:
\begin{align}
 \!\not \epsilon^- (p_i,r) = \,&\frac{\sqrt{2}}{[ri]} (  \vert i \rangle [ r \vert + \vert r ] \langle i \vert  )  ,\label{e1}\\ 
 \! \not \epsilon^+ (p_i,r)  =\,& \frac{\sqrt{2}}{\langle ri \rangle} (  \vert i ] \langle r \vert + \vert r \rangle [ i \vert  )    ,\label{e0}\\
    \epsilon^+(p_i, r) \cdot p_j = \,&\frac{ [ij] \langle jr \rangle  }{ \sqrt{2} \langle ri \rangle  },\label{epj0}\\
 \epsilon^-(p_i,r) \cdot p_j =\, &\frac{ \langle ij \rangle  [jr]     }{\sqrt{2} [ir] }   ,\label{epj1}\\
 \epsilon^-(p_i,r) \cdot \epsilon^+(p_j,q) =\,& \frac{ \langle  iq  \rangle [jr]  }{ 
[ir] \langle   qj    \rangle  } ,  \label{e1e0}
\end{align}
here $i$ and $j$ are the momenta of the particles such as gluons, $r$ and $q$ are their reference momenta which are arbitrary  and $p_i \neq r$. Polarization vectors must satisfy the following on-shell conditions \cite{Carrasco:2015iwa},
\begin{equation}\label{OSR}
    \epsilon^{\pm}(p_i,r) \cdot p_i = 0, \quad \epsilon^{\pm} (p_i,r) \cdot \epsilon^{\pm}(p_j,q) = 0.
\end{equation}
In a supergravity theory, the consequence of KLT relations is that polarization tensors of the gravitino field can be written in terms of gluino and gluons as \cite{Bjerrum-Bohr:2010eeh}: 
\begin{align}
\widetilde{\epsilon}^{+ \mu}_{\widetilde{h}} (p, q)    =  \,&\widetilde{\epsilon}^+_{\widetilde{g}}(p) \times \epsilon^{+ \mu}_g (p,q) = [p \vert \times \epsilon^{+ \mu}_g (p,q),\label{h1}\\
\widetilde{\epsilon}^{- \mu}_{\widetilde{h}} (p, q)    =  \,&\widetilde{\epsilon}^-_{\widetilde{g}}(p) \times \epsilon^{- \mu}_g (p,q) = \langle p \vert \times \epsilon^{- \mu}_g (p,q),\label{h2}\\
\epsilon^{+ \mu}_{\widetilde{h}} (p, q)    =  \,&\epsilon^+_{\widetilde{g}}(p) \times \epsilon^{+ \mu}_g (p,q) =  \vert  p ]  \times \epsilon^{+ \mu}_g (p,q),\label{h3}\\
\epsilon^{- \mu}_{\widetilde{h}} (p, q)    = \,& \epsilon^-_{\widetilde{g}}(p) \times \epsilon^{- \mu}_g (p,q) =  \vert p \rangle   \times \epsilon^{- \mu}_g (p,q).\label{h4}
\end{align}
The same is applicable for the polarization tensor of the graviton field which in  terms of the gluon field is expressed as:
\begin{equation}\label{h00}
   ( \epsilon^{++}_h)^{\mu \nu}(p,q) = \epsilon^{+ \mu}_g(p,q) \times \epsilon^{+ \nu}_g (p,q),
\end{equation}
\begin{equation}\label{h11}
   (\epsilon^{--}_h)^{\mu \nu} (p,q) = \epsilon^{- \mu}_g (p,q) \times \epsilon^{- \nu}_g (p,q),
\end{equation}
where $q$ means reference momentum. Furthermore, we can express contraction between gravitino and graviton fields in terms of Weyl spinors through equations (\ref{e1}, \ref{e1e0}) due to the presence of KLT relations. On the other hand, according to \cite{Schwartz:2014sze}, partial amplitudes for four gluons in their corresponding channels are: 
\begin{align}\label{KLTgluons}
   & s_{12}: \quad A_4(1^-_g, 2^-_g, 3^+_g, 4^+_g) = \frac{\langle 12 \rangle^4}{\langle 12 \rangle \langle 23 \rangle \langle 34 \rangle \langle 41 \rangle},\\ 
   & s_{13}: \quad A_4 (1^-_g, 3^+_g, 2^-_g, 4^+_g) = \frac{\langle 21 \rangle^4}{ \langle 14 \rangle \langle 42 \rangle \langle 23 \rangle \langle 31 \rangle}, \\ 
& s_{14}: \quad A_4 (1^-_g, 4^+_g , 2^-_g, 3^+_g)   = \quad 0.
\end{align}
Finally, partial amplitudes for two gluinos and two gluons are \cite{ReyesPerez:2021amr}:
\begin{align}\label{KLTgluinos}
   & s_{12}: \quad A_4(1^-_{\widetilde{g}}, 2^-_g, 3^+_{\widetilde{g}}, 4^+_g) = \frac{\langle 12 \rangle^3}{\langle 12 \rangle \langle 34 \rangle  \langle 41 \rangle},\\ 
   & s_{13}: \quad A_4 (1^-_{\widetilde{g}}, 3^+_{\widetilde{g}}, 2^-_g, 4^+_g) = \frac{\langle 21 \rangle^3}{ \langle 14 \rangle \langle 42 \rangle  \langle 31 \rangle}, \\ 
& s_{14}: \quad A_4 (1^-_{\widetilde{g}}, 4^+_g, 2^-_g, 3^+_{\widetilde{g}})   = \quad 0.
\end{align}

\section{Supergravity N=1}\label{Sugradetails}


In this appendix, we present the invariance of the $\mN=1$ globally supersymmetric lagrangian for the graviton and its spin $\frac{3}{2}$ partner -- the gravitino, under supersymmetry transformations \cite{Fierz:1939ix,Rarita:1941mf}. We construct the supersymmetry transformation in terms of two-component Weyl spinor formalism to be able to connect with the calculations  done in section (\ref{section4}); this is useful for deriving Feynmann rules directly using two-component spinors \cite{Dreiner:2008tw}.

Supergravity is primarily known as a theory with local supersymmetry where the number $\mN=1$ means that the chiral supermultiplet will contain one superpartner of spin-2 graviton. We consider pure supergravity with zero coupling to matter fields as well as the presence of both graviton and gravitino. We leave the problem of constructing a locally supersymmetric Langragian for the supergravity multiplet--written in Weyl spinor basis for future work. The on-shell globally supersymmetric pure supergravity (no matter coupling) Lagrangian can be ideally described  as a summation of quadratic kinetic terms of both the gravitino and graviton. More specifically, it contains two bosonic degrees of freedom of the graviton and two fermionic degrees of freedom of the massless Weyl vector spinor. The kinetic term of the gravitino is written in terms of massless Dirac vector spinor field $\Psi_{\mu}$ and the graviton in terms of a traceless symmetric tensor $g_{\mu\nu}$. Let us now check the invariance of the total lagrangian against the global supersymmetric transformation individually for both graviton and gravitino.



\subsection{Rarita-Schwinger in two-component formalism}\label{appA}
The Rarita-Schwinger action of \cite{Rarita:1941mf,Bailin:1994qt} accurately describes the kinetic term of a gravitino written in terms of a Dirac spinor:
\begin{equation}
    S_{RS} = - \frac{1}{2} \int d^4 x \epsilon^{\mu \nu \rho \sigma} \overline{\Psi}_{\mu} \gamma_5 \gamma_{\nu} \partial_{\rho} \Psi_{\sigma},
\end{equation}
where $\epsilon^{\epsilon\nu\rho\sigma}$ is the totally antisymmetric Levi-Civita symbol with $\epsilon^{0123}=1$. The Dirac spinor is
\begin{equation}\label{weylspinor}
  \Psi_{\mu} =	\left(
	\begin{array}{c}
		(\psi_{\mu})_a \\
		(\widetilde{\chi}_{\mu})^{\dot{a}}
	\end{array} \right), \quad
 \overline{\Psi}_{\mu} = \Psi^{\dagger}_{\mu} \gamma^0.
\end{equation}
where $\psi_{\mu},\tilde{\chi}_{\mu}$ are the Weyl spinors of opposite chirality. Some matrix definitions in Chiral representation are the following:

\begin{equation}\label{gammaweyl}
    \gamma_{\nu} = 	\left(
	\begin{array}{cc}
		0 &  (\sigma_{\nu})_{a \dot{a}} \\
		(\Bar{\sigma}_{\nu})^{\dot{a} a} &  0
	\end{array} \right), \quad
\gamma_5= 	\left(
	\begin{array}{cc}
		-1 &  0 \\
		 0 &  1
	\end{array} \right), \quad
 \gamma^0 =	\left(
	\begin{array}{cc}
		0 &  1 \\
		1 &  0
	\end{array} \right).
\end{equation}
These Weyl spinors can be written as
\begin{equation}
    \widetilde{\psi}_{\dot{a}} \equiv (\psi_a)^*, \quad \chi^{a} \equiv (\widetilde{\chi}^{\dot{a}})^*,
\end{equation}
which helps us to express the adjoint Dirac spinor as
\begin{equation}
    \Psi^{\dagger}_{\mu} = ( (\widetilde{\psi}_{\mu})_{\dot{a}} , (\chi_{\mu})^{a}).
\end{equation}
Using the above definitions, the Rarita-Schwinger action in terms of Weyl spinors becomes:
\begin{align}\label{WeylRS}
 S_{RS} &= - \frac{1}{2} \int d^4x \epsilon^{\mu \nu \rho \sigma} (   (\widetilde{\psi}_{\mu})_{\dot{a}} (\Bar{\sigma}_{\nu})^{\dot{a} a} \partial_{\rho} (\psi_{\sigma})_{a} - (\chi_{\mu})^{a}  (\sigma_{\nu})_{a \dot{a}} \partial_{\rho} (\widetilde{\chi}_{\sigma})^{\dot{a}}    )       \nonumber  \\
  & = - \frac{1}{2} \int d^4x \epsilon^{\mu \nu \rho \sigma} (   \widetilde{\psi}_{\mu} \Bar{\sigma}_{\nu} \partial_{\rho} \psi_{\sigma} - \chi_{\mu} \sigma_{\nu} \partial_{\rho} \widetilde{\chi}_{\sigma}    ).  
\end{align}
The equations of motion associated with the two fermionic components of the gravitino $\tilde{\psi_{\mu}},\chi_{\mu}$ of (\ref{WeylRS}) are
\begin{equation}\label{EM1}
    \epsilon^{\mu \nu \rho \sigma} (\Bar{\sigma}_{\nu})^{\dot{a} a} \partial_{\rho} (\psi_{\sigma})_a = 0,
\end{equation}
\begin{equation}\label{EM2}
    \epsilon^{\mu \nu \rho \sigma} (\sigma_{\nu})_{a \dot{a}} \partial_{\rho} (\widetilde{\chi}_{\sigma})^{\dot{a}} = 0.
\end{equation}
To obtain the correct kinetic term which is quadratic in $g_{\mu\nu}$, we need to make use of linearised Einstein gravity where $g_{\mu\nu}=\eta_{\mu\nu}+\kappa h_{\mu\nu}$. 
Here, $\eta_{\mu\nu}$ is the Minkowski metric and $\kappa^2=8\pi G_N$, where we take the same convention as in ref.\cite{Bailin:1994qt}, that is $\kappa^2=1$.
The factor $\kappa$ helps to keep the dimension of $h_{\mu\nu}$ correct that is mass dimension 1. One can expand the Einstein equation to linear order in $h_{\mu\nu}$ and obtain \eqref{3}, however, in this subsection we will only consider the RS-action and its supersymmetric invariance. Next, let us consider a global supersymmetry transformation of the form $e^{i\xi Q}$ with $\xi$ being the Majorana spinor parameter and $Q$ the Majorana spinor supersymmetry generator. The global supersymmetric transformations in two components are:
\begin{equation}\label{psia}
    \delta_{\xi} (\psi_{\sigma})_a = - i (\sigma^{\theta \tau})_a^b \partial_{\theta} h_{\tau \sigma} \xi_b,
\end{equation}
\begin{equation}\label{chiadot}
   \delta_{\xi} (\widetilde{\chi}_{\sigma})^{\dot{a}} = -i (\Bar{\sigma}^{\theta \tau})^{\dot{a}}_{\dot{b}} \partial_{\theta} h_{\tau \sigma} \widetilde{\xi}^{\dot{b}}. 
\end{equation}
Let us now check the invariance of (\ref{WeylRS}) under the supersymmetry transformations of (\ref{psia}) and (\ref{chiadot}):
\begin{equation}
   \delta_{\xi} S_{RS} = - \frac{1}{2} \int d^4x \epsilon^{\mu \nu \rho \sigma} \left[  
   ( \delta_{\xi} \widetilde{\psi}_{\mu}  ) \Bar{\sigma}_{\nu} \partial_{\rho}\psi_{\sigma} + \widetilde{\psi}_{\mu} \Bar{\sigma}_{\nu} \partial_{\rho}(\delta_{\xi} \psi_{\sigma}) -  (\delta_{\xi} \chi_{\mu}) \sigma_{\nu} \partial_{\rho} \widetilde{\chi}_{\sigma} - \chi_{\mu} \sigma_{\nu} \partial_{\rho}(\delta_{\xi} \widetilde{\chi}_{\sigma} )    \right].
\end{equation}
Imposing the equations of motion (\ref{EM1}) and (\ref{EM2}), we find that the both first and third terms vanish and substituting the global supersymmetry transformations, we are left with the following:
\begin{equation}
  \delta_{\xi} S_{RS} = - \frac{1}{2} \int d^4x \epsilon^{\mu \nu \rho \sigma} \left[         
    (\widetilde{\psi}_{\mu})_{\dot{a}}(\Bar{\sigma}_{\nu})^{\dot{a} a} (\sigma^{\theta \tau})_a^b \xi_b - (\chi_{\mu})^a(\sigma_{\nu})_{a \dot{a}} (\Bar{\sigma}^{\theta \tau})^{\dot{a}}_{\dot{b}} \widetilde{\xi}^{\dot{b}}  \right]  \partial_{\rho} \partial_{\theta} h_{\tau \sigma}. 
\end{equation}
We define $(A^{\theta \tau})_a = (\sigma^{\theta \tau})_a^b \xi_b$ and $(\widetilde{A}^{\theta \tau})^{\dot{a}} = (\Bar{\sigma}^{\theta \tau})^{\dot{a}}_{\dot{b}} \widetilde{\xi}^{\dot{b}}$. Applying the following property of spinors, $\widetilde{\chi}_{\dot{a}} (\Bar{\sigma}^{\mu})^{\dot{a} a} \psi_a = - \psi^a (\sigma^{\mu})_{a \dot{a}} \widetilde{\chi}^{\dot{a}}$, we get:
\begin{align}\label{SRSA}
    \delta_{\xi}  S_{RS} &= - \frac{1}{2} \int d^4x \epsilon^{\mu \nu \rho \sigma} \left[  
   (\widetilde{A}^{\theta \tau})_{\dot{a}} (\Bar{\sigma}_{\nu})^{\dot{a} a} (\chi_{\mu})_{a} - (A^{\theta \tau})^a (\sigma_{\nu})_{a \dot{a}} (\widetilde{\psi}_{\mu})^{\dot{a}}              \right] \partial_{\rho} \partial_{\theta} h_{\tau \sigma} \nonumber\\ 
   & = - \frac{1}{2} \int d^4x \epsilon^{\mu \nu \rho \sigma} \left[  \widetilde{B}^{\theta \tau}_{\nu} \chi_{\mu} - B^{\theta \tau}_{\nu} \widetilde{\psi}_{\mu}   \right] \partial_{\rho} \partial_{\theta} h_{\tau \sigma}.
\end{align}
 We further define  $(B^{\theta \tau}_{\nu})_{\dot{a}} = (A^{\theta \tau})^a (\sigma_{\nu})_{a \dot{a}}$ and $(\widetilde{B}^{\theta \tau}_{\nu})^a = \widetilde{A}^{\theta \tau}_{\dot{a}} (\Bar{\sigma}_{\nu})^{\dot{a} a}$. Next, we apply the product derivative rule  for simplifications:
\begin{align}
  \delta_{\xi}  S_{RS} =\,& - \frac{1}{2} \int d^4x \epsilon^{\mu \nu \rho \sigma} \bigg[                \partial_{\rho} (\widetilde{B}^{\theta \tau}_{\nu} \chi_{\mu} \partial_{\theta} h_{\tau \sigma}) - \partial_{\rho} (  B^{\theta \tau}_{\nu}  \widetilde{\psi}_{\mu}   \partial_{\theta} h_{\tau \sigma})\nonumber\\
  \,& + B^{\theta \tau}_{\nu} \partial_{\rho} \widetilde{\psi}_{\mu} \partial_{\theta} h_{\tau \sigma} - \widetilde{B}^{\theta \tau}_{\nu} \partial_{\rho} \chi_{\mu} \partial_{\theta} h_{\tau \sigma}     \bigg].  
\end{align}
Third and fourth terms are zero as they are the equations of motion
\begin{align}
    A^{\theta \tau} \epsilon^{\mu \nu \rho \sigma} \sigma_{\nu} \partial_{\rho} \widetilde{\psi}_{\mu} \partial_{\theta} h_{\tau \theta} &= 0,\label{eqmRS1}\\
    \widetilde{A}^{\theta \tau} \epsilon^{\mu \nu \rho \sigma}  \Bar{\sigma}_{\nu} \partial_{\rho} \chi_{\mu} \partial_{\theta} h_{\tau \sigma}& = 0.\label{EqmRS2}
\end{align}
Substituting the above equation of motions, we finally arrive at a total derivative 
\begin{align}
 \delta_{\xi}  S_{RS} &= - \frac{1}{2} \int d^4x \epsilon^{\mu \nu \rho \sigma}   \partial_{\rho}    \left[     \widetilde{A}^{\theta \tau}_{\dot{a}}(\Bar{\sigma}_{\nu})^{\dot{a} a} (\chi_{\mu})_a \partial_{\rho} h_{\tau \sigma} - (A^{\theta \tau})^a (\sigma_{\nu})_{a \dot{a}} (\widetilde{\psi}_{\mu})^{\dot{a}} \partial_{\theta}  h_{\tau \sigma}    \right] \nonumber\\ 
 & =   - \frac{1}{2} \int d^4x \epsilon^{\mu \nu \rho \sigma}   \partial_{\rho} \left[  \partial_{\theta} h_{\tau \sigma} ( \widetilde{\psi}_{\mu} \Bar{\sigma}_{\nu} \sigma^{\theta \tau} \xi - \chi_{\mu} \sigma_{\nu} \Bar{\sigma}^{\theta \tau}  \widetilde{\xi}   )    \right].
\end{align}
This shows that the Rarita Schwinger action  in terms of Weyl spinors (two component spinors) is invariant under global SUSY transformations. Now, the second task is to demonstrate that the Rarita-Schwinger action is invariant under gauge transformations (diffeomorphisms). The gauge transformation corresponding to the Rarita-Schwinger action in terms of dirac spinors is given by
\begin{equation} \label{9}
    \delta_{\eta}  \Psi_{\mu}= \partial_{\mu} \eta,
\end{equation}
where $\eta$ is a fermionic parameter, which we may write in terms of Weyl spinors as follows:
\begin{equation}\label{weylspinoreta}
  \eta = \left(
	\begin{array}{c}
		(\zeta)_a \\
		(\widetilde{\lambda})^{\dot{a}}
\end{array} \right).
\end{equation}
The transformation  of (\ref{9}) can be written in two components as follows
\begin{equation}\label{transRS1}
    \delta_\eta(\psi_\mu)=\partial_\mu(\zeta)_a,
\end{equation}
\begin{equation}\label{transRS2}
\delta_\eta(\widetilde{\chi}_\mu)=\partial_\mu(\lambda)^{\dot{a}}.
\end{equation}
The aim is to verify that these gauge transformations leave the action (\ref{WeylRS}) invariant. The variation of the Rarita-Schwinger lagrangian $\mathcal{L}_{RS}$ under the transformations (\ref{transRS1})-(\ref{transRS2}) is given as follows
\begin{align}
    \delta_{\eta} \mathcal{L}_{RS} & = -\frac{1}{2}\epsilon^{\mu \nu \rho \sigma}\biggl\{[\delta_\eta(\widetilde{\psi}_\mu)_{\dot{a}}](\bar{\sigma}_\nu)^{\dot{a}a}\partial_\rho(\psi_\sigma)_a+(\widetilde{\psi}_\mu)_{\dot{a}}(\bar{\sigma})^{\dot{a}a}\partial_\rho[\delta_\eta(\psi_\sigma)_a]\notag\\
    &\quad-[\delta_\eta(\chi_\mu)^a](\sigma_\nu)_{a\dot{a}}\partial_\rho(\widetilde{\chi}_\sigma)^{\dot{a}}-(\chi_\mu)^a(\sigma_\nu)_{a\dot{a}}\partial_\rho[\delta_\eta(\widetilde{\chi}_\sigma)^{\dot{a}}]
    \biggr\}\notag\\
    & = -\frac{1}{2} \epsilon^{\mu \nu \rho \sigma}\biggl\{ (\widetilde{\psi}_\mu)_{\dot{a}}(\bar{\sigma})^{\dot{a}a}\partial_\rho[\delta_\eta(\psi_\sigma)_a]-(\chi_\mu)^a(\sigma_\nu)_{a\dot{a}}\partial_\rho[\delta_\eta(\widetilde{\chi}_\sigma)^{\dot{a}}]\biggr\}\notag\\
    & = -\frac{1}{2} \epsilon^{\mu \nu \rho \sigma}\biggl\{ 
    \partial_\rho\{(\widetilde{\psi}_\mu)_{\dot{a}}(\bar{\sigma}_\nu)^{\dot{a}a}[\delta_\eta(\psi_\sigma)_a]\}
    -\partial_\rho(\widetilde{\psi}_\mu)_{\dot{a}}(\bar{\sigma}_\nu)^{\dot{a}a}[\delta_\eta(\psi_\sigma)_a] \notag\\
    &\quad-\partial_\rho\{(\chi_\mu)^a(\sigma_\nu)_{a\dot{a}}[\delta_\eta(\widetilde{\chi}_\sigma)^{\dot{a}}]\}+\partial_\rho(\chi_\mu)^a(\sigma_\nu)_{a\dot{a}}[\delta_\eta(\widetilde{\chi}_\sigma)^{\dot{a}}]
    \biggr\}\notag\\
    & = -\frac{1}{2} \epsilon^{\mu \nu \rho \sigma}\partial_\rho\biggl\{(\widetilde{\psi}_\mu)_{\dot{a}}(\bar{\sigma}_\nu)^{\dot{a}a}[\delta_\eta(\psi_\sigma)_a]-(\chi_\mu)^a(\sigma_\nu)_{a\dot{a}}[\delta_\eta(\widetilde{\chi}_\sigma)^{\dot{a}}]\biggr\} \notag\\
    & = -\frac{1}{2} \epsilon^{\mu \nu \rho \sigma}\partial_\rho\biggl\{\widetilde{\psi}_\mu\bar{\sigma}_\nu(\delta_\eta\psi_\sigma)-\chi_\mu\sigma_\nu(\delta_\eta\widetilde{\chi}_\sigma)\biggr\}\notag\\
    & = -\frac{1}{2} \epsilon^{\mu \nu \rho \sigma}\partial_\rho\biggl\{\widetilde{\psi}_\mu\bar{\sigma}_\nu\partial_\sigma\zeta-\chi_\mu\sigma_\nu\partial_\sigma\widetilde{\lambda}\biggr\}\notag
\end{align}
where in the second line, we used the field equations (\ref{eqmRS1})-(\ref{EqmRS2}), and in the last line, the Leibniz rule was applied. Also, it can be noted that
\begin{align}
    \epsilon^{\mu \nu \rho \sigma}  \gamma_5 \gamma_{\nu}(\partial_\rho\bar{\Psi}_\mu)& =\epsilon^{\sigma \nu \rho \mu}  \gamma_5 \gamma_{\nu} (\partial_\rho\bar{\Psi}_\sigma)\notag\\
    & =\epsilon^{\mu \nu \rho \sigma}  \gamma_5 \gamma_{\nu}(\partial_\rho\bar{\Psi}_\sigma)\notag\\
    & =0,
\end{align}
where, in the last row, we again used (\ref{eqmRS1})-(\ref{EqmRS2}). Thus, the variation of the Rarita-Schwinger Lagrangian under gauge transformation (\ref{9}) is
\begin{equation}
    \delta_{\eta} \mathcal{L}_{RS}=-\frac{1}{2}\epsilon^{\mu\nu\rho\sigma}\gamma_5\gamma_{\nu}\partial_\rho[\bar{\Psi}_\mu(\partial_\sigma\eta)],
\end{equation}
and being a total derivative, this implies that $\delta_{\eta} \mathcal{S}_{RS}=0$.

\subsection{Einstein-Hilbert in two-components formalism}\label{appB}

In this subsection, we look at the kinetic term of the graviton and check its invariance. The linearised Einstein-Hilbert action by expanding the metric in  linear order in $h_{\mu\nu}$ is given by \cite{Bailin:1994qt}:

\begin{equation} \label{3}
	S_{EH} = - \frac{1}{2} \int d^4 x (   R^L_{\mu \nu} - \frac{1}{2}  \eta_{\mu \nu} R^L         ) h^{\mu \nu}.
\end{equation}
The corresponding linearized Einstein field equation in vacuum is
\begin{equation} \label{4}
	R^L_{\mu \nu} = 0,
\end{equation}
where the linearized Ricci tensor $R^L_{\mu \nu}$ is expressed as,
\begin{equation}
	R^L_{\mu \nu} = \frac{1}{2}  (    \partial^{\lambda} \partial_{\mu} h_{\lambda \nu} + \partial^{\lambda} \partial_{\nu}  h_{\lambda \mu} - \Box h_{\mu \nu} - \partial_{\mu} \partial_{\nu} h                     ),
\end{equation}
and the Ricci scalar $R^L$ is
\begin{equation}
	R^L \equiv \eta^{\mu\nu} R^L_{\mu\nu}.
\end{equation}
The global supersymmetry transformation for the $h_{\mu \nu}$ is given by,
\begin{equation} \label{12}
\begin{split}
    \delta_{\xi} h_{\mu \nu} & =-\frac{i}{2}\bar{\xi}\left(\gamma_\mu\Psi_\nu+\gamma_\nu\Psi_\mu \right) \\
    & = -\frac{i}{2}\biggl\{ \xi^a((\sigma_\mu)_{a\dot{a}}\tilde{\psi}_\nu^{\dot{a}}+(\sigma_\nu)_{a\dot{a}}\tilde{\psi}_\mu^{\dot{a}})+\tilde{\xi}_{\dot{a}}((\bar{\sigma}_\mu)^{\dot{a}a}\psi_{\nu a}+(\bar{\sigma}_\nu)^{\dot{a}a}\psi_{\mu a})\biggr\} \\
    & = -\frac{i}{2}\biggl\{ \xi(\sigma_\mu\tilde{\psi}_\nu+\sigma_\nu\tilde{\psi}_\mu)+\tilde{\xi}(\bar{\sigma}_\mu\psi_\nu+\bar{\sigma}_\nu\psi_\mu)\biggr\},
\end{split}
\end{equation}
which is written in terms of Weyl spinors using (\ref{weylspinor}) and (\ref{gammaweyl}). Before showing how the supersymmetry transformation (\ref{12}) leaves the action (\ref{3}) invariant, let us make a remark on the free spinor field $\Psi^{\mu}$. A spinor field with an additional vector index belongs to the representation of the Lorentz homogeneous group $[ (\frac{1}{2}, 0     ) +  (  0, \frac{1}{2} )   ] \times ( \frac{1}{2}, \frac{1}{2}  )$. In order to isolate the part $ (1, \frac{1}{2} )   +   (\frac{1}{2}, 1  ) $ of the free field, the following irreducibility conditions should be imposed:
\begin{equation}\label{irrcon1}
	\gamma_{\mu} \Psi^{\mu} = 0,
\end{equation}
\begin{equation}\label{irrcon2}
	\partial_{\mu} \Psi^{\mu} = 0,
\end{equation}
as well as the  Dirac equation
\begin{equation}\label{eqdirac}
	\! \not \partial \Psi^{\mu} = \gamma^\nu\partial_\nu\Psi^\mu= 0,
\end{equation}
these conditions may also be written in terms of Weyl spinors. As a result, we obtain two equations for every irreduciability condition
\begin{align}
    \sigma^\mu\tilde{\psi}_\mu&=0, \label{6}\\
    \bar{\sigma}^\mu\psi_\mu&=0 \label{61},\\
    \partial_\mu\tilde{\psi}^\mu&=0, \label{7}\\
    \partial_\mu\psi^\mu&=0, \label{71}\\
    \partial_\nu\sigma^\nu\tilde{\psi}^\mu&=0, \label{8}\\
    \partial_\nu\bar{\sigma}^\nu\psi^\mu&=0, \label{81}
\end{align}
where (\ref{6}) and (\ref{61}), (\ref{7}) and (\ref{71}), and (\ref{8}) and (\ref{81}), refer to (\ref{irrcon1}), (\ref{irrcon2}), and (\ref{eqdirac}), respectively, written in two components. This conditions will be useful when showing that the SUGRA lagrangian is invariant under SUSY transformations.

The previous conditions may be derived by considering the little group, in particular, by constructing the Pauli-Lubanski and momentum operators, and from them building up the projectors in different subspaces with spin $3/2$ and spin $1/2$ \cite{Kirchbach:2002nu,Napsuciale:2006wr}. Also, they provide a useful computational advantage in proving explicitly that the SUGRA action is indeed invariant under SUSY transformations.
We proceed by checking that the linearized Einstein-Hilbert action(\ref{3}) is actually invariant under the supersymmetry tranformation (\ref{12}):
\begin{equation}
    \begin{split} \label{13}
        \delta_{\xi} \mathcal{L}_{EH}&= -\frac{1}{2}\delta_\xi\bigl\{(R^L_{\mu \nu} - \frac{1}{2}  \eta_{\mu \nu} R^L         ) h^{\mu \nu}\bigr\} \\
        &= \frac{1}{4}\delta_\xi\left(hR^L\right)-\frac{1}{2}\delta_\xi\left(h^{\mu\nu}R^L_{\mu\nu}\right).
    \end{split}
\end{equation}
First, we begin by computing every term in (\ref{13}) independently and for the first term we get
\begin{equation} \label{18}
    \begin{split}
        \delta_\xi\left(hR^L\right) &= (\delta_\xi h)R^L+h(\delta_\xi R^L) \\
        &= h[\partial^\mu\partial_\gamma(\delta_\xi h^\gamma_{\;\;\mu})-\Box(\delta_\xi h)]\\
        &=0.
    \end{split}
\end{equation}
Notice that $R^L=0$ comes as a result from the field equations (\ref{4}), and that any term of the form $\partial_\mu(\delta_\xi h^\mu_\nu)$ or $\delta_{\xi}h$ vanishes since this lead to the irreducibility conditions (\ref{7})-(\ref{81}).
Secondly, we compute the other term in (\ref{13}), and obtain the result
\begin{align} \label{ecterm2}
        \delta_\xi\left(h^{\mu\nu}R^L_{\mu\nu}\right)&= \frac{1}{2}\delta_\xi[h^{\mu\nu}(\partial_\mu\partial_\gamma h^\gamma_{\;\;\nu}+\partial_\nu\partial_\gamma h^\gamma_{\;\;\mu}-\partial_\mu\partial_\nu h-\Box h_{\mu\nu})]\nonumber\\
        &= \frac{1}{2}(\delta_\xi h^{\mu\nu})(\partial_\mu\partial_\gamma h^\gamma_{\;\;\nu}+\partial_\nu\partial_\gamma h^\gamma_{\;\;\mu}-\partial_\mu\partial_\nu h-\Box h_{\mu\nu})-\frac{1}{2}h^{\mu\nu}\Box(\delta_\xi h_{\mu\nu})\nonumber\\
        &= \frac{1}{2}(\delta_\xi h^{\mu\nu})\partial_\mu(2\partial_\gamma h^\gamma_{\;\;\nu}-\partial_\nu h)-\frac{1}{2}(\delta_\xi h^{\mu\nu})\Box h_{\mu\nu}-\frac{1}{2}h^{\mu\nu}\Box(\delta_\xi h_{\mu\nu})\nonumber\\
        &= \partial_\mu\bigl\{(\delta_\xi h^{\mu\nu})(\partial_\gamma h^\gamma_{\;\;\nu}-\frac{1}{2}\partial_\nu h)\bigr\}-\frac{1}{2}(\delta_\xi h^{\mu\nu})\Box h_{\mu\nu}-\frac{1}{2}h^{\mu\nu}\Box(\delta_\xi h_{\mu\nu}),
\end{align}
since, as before, the terms of the form $\partial_\mu(\delta_\xi h^\mu_\nu)$ and $\delta_\xi h$ vanish. Note that,
\begin{equation}
    \begin{split}
        \Box[h_{\mu\nu}(\delta_\xi h^{\mu\nu})] &= \partial^\lambda[(\partial_\lambda h_{\mu\nu})(\delta_\xi h^{\mu\nu})+h_{\mu\nu}(\partial_\lambda\delta_\xi h^{\mu\nu})]\\
        &= (\delta_\xi h^{\mu\nu})\Box h_{\mu\nu}+2(\partial_\lambda h_{\mu\nu})(\partial^\lambda\delta_\xi h^{\mu\nu})+\Box(\delta_\xi h^{\mu\nu}),
    \end{split}
\end{equation}
and by making some rearrangements in the tensor indices and substituting back into (\ref{ecterm2}), we can rewrite \eqref{ecterm2}  as,
\begin{equation} \label{14}
    \begin{split}
        \delta_\xi\left(h^{\mu\nu}R^L_{\mu\nu}\right) &= \partial^\lambda\bigl\{(\delta_\xi h^\nu_{\;\;\lambda})(\partial_\gamma h^\gamma_{\;\;\nu}-\frac{1}{2}\partial_\nu h)-\frac{1}{2}\partial_\lambda[h_{\mu\nu}(\delta_\xi h^{\mu\nu})]\bigr\}+(\partial_\lambda h_{\mu\nu})(\partial^\lambda\delta_\xi h^{\mu\nu}).
    \end{split}
\end{equation}
Now, let us focus on the last term which is outside the total derivative in \eqref{14}. After some algebraic manipulation, we obtain
\begin{equation} \label{15}
    \begin{split}
        \partial^\lambda[(\delta_\xi h_{\mu\nu})(\partial_\lambda h^{\mu\nu})] &= (\partial^\lambda \delta_\xi h_{\mu\nu})(\partial_\lambda h^{\mu\nu})+(\delta_\xi h_{\mu\nu})\Box h^{\mu\nu}\\
        &=(\partial^\lambda \delta_\xi h_{\mu\nu})(\partial_\lambda h^{\mu\nu})+(\delta_\xi h_{\mu\nu})(\partial^\mu\partial_\gamma h^{\gamma\nu}+\partial^\nu\partial_\gamma h^{\gamma\mu}-\partial^\mu\partial^\nu h),
    \end{split}
\end{equation}
where we made use of the field equation (\ref{4}) written in explicit form. Nevertheless, notice that the following terms are equivalent to the total derivative on the left side of (\ref{ex1}) and (\ref{ex2}),
\begin{equation}\label{ex1}
    \partial^\mu[(\delta_\xi h_{\mu\nu})(\partial_\gamma h^{\gamma\nu})]=(\delta_\xi h_{\mu\nu})(\partial^\nu\partial_\gamma h^{\gamma\mu}),
\end{equation}
\begin{equation}\label{ex2}
        \partial^\nu[(\delta_\xi h_{\mu\nu})(\partial^\mu h)]=(\delta_\xi h_{\mu\nu})(\partial^\mu\partial_\nu h),
\end{equation}
since in the first term they both have the form $\partial^\nu \delta_\xi h_{\mu\nu}$, the first terms in each equation vanish.

Subsequently, we may rewrite equation (\ref{15}) as
\begin{equation} \label{16}
    \begin{split}
        (\partial^\lambda \delta_\xi h_{\mu\nu})(\partial_\lambda h^{\mu\nu})&= \partial^\lambda[(\delta_\xi h_{\mu\nu})(\partial_\lambda h^{\mu\nu})]-\partial^\mu[(\delta_\xi h_{\mu\nu})(2\partial_\gamma h^{\gamma\nu}-\partial^\nu h)] \\
        &= \partial^\lambda[(\delta_\xi h_{\mu\nu})(\partial_\lambda h^{\mu\nu})-(\delta_\xi h_{\lambda\nu})(2\partial_\gamma h^{\gamma\nu}-\partial^\nu h)].
    \end{split}
\end{equation}
After making some rearrangements in the tensor indices and substituting (\ref{16}) into (\ref{14}), we find
\begin{equation} \label{17}
    \begin{split}
        \delta_\xi \left(h^{\mu\nu}R^L_{\mu\nu}\right)&=\partial^\lambda\bigl\{(\delta_\xi h_{\lambda\nu})(\frac{1}{2}\partial^\nu h-\partial_\gamma h^{\gamma\nu})+(\delta_\xi h_{\mu\nu})(\partial_\lambda h^{\mu\nu})-\frac{1}{2}\partial_\lambda[h_{\mu\nu}(\delta_\xi h^{\mu\nu})] \bigr\}.\\
    \end{split}
\end{equation}
To conclude, by substituting (\ref{18}) and (\ref{17}) in (\ref{13}), we find that the change in the lagrangian as

\begin{equation}
    \begin{split} \label{final}
    \delta_{\xi} \mathcal{L}_{EH} & = \partial^\lambda\bigl\{(\delta_\xi h_{\lambda\nu})(\frac{1}{2}\partial_\gamma h^{\gamma\nu}-\frac{1}{4}\partial^\nu h)+\frac{1}{4}\partial_\lambda[h_{\mu\nu}(\delta_\xi h^{\mu\nu})]-\frac{1}{2}(\delta_\xi h_{\mu\nu})(\partial_\lambda h^{\mu\nu})\bigr\} \\
    &= \partial^\lambda\bigl\{(\delta_\xi h_{\lambda\nu})(\frac{1}{2}\partial_\gamma h^{\gamma\nu}-\frac{1}{4}\partial^\nu h)+\frac{1}{4}h^{\mu\nu}(\partial_\lambda\delta_\xi h_{\mu\nu})-\frac{1}{4}(\partial_\lambda h^{\mu\nu})(\delta_\xi h_{\mu\nu})\bigr\}
\end{split}
\end{equation}

Finally, by substituting the supersymmetry transformation (\ref{12}) into equation (\ref{final}), we obtain  the corresponding variation in the Einstein-Hilbert lagrangian which is equivalent to the following total derivative:
\begin{align}
    \delta_{\xi} \mathcal{L}_{EH}  =\,& \frac{i}{4}\partial^\lambda\biggl\{ \biggr(\frac{1}{2}\partial^\nu h-\partial_\gamma h^{\gamma\nu}\biggr)[\xi(\sigma_\lambda \tilde{\psi}_\nu+\sigma_\nu\tilde{\psi}_\lambda)+\tilde{\xi}(\bar{\sigma}_\lambda+\bar{\sigma}_\nu\psi_\lambda)] \nonumber\\
    & \quad\quad+\partial_\lambda h^{\mu\nu}(\xi\sigma_\mu\tilde{\psi}_\nu+\tilde{\xi}\bar{\sigma}_\mu\psi_\nu)-h^{\mu\nu}\partial_\lambda(\xi\sigma_\mu\tilde{\psi}_\nu+\tilde{\xi}\bar{\sigma}_\mu\psi_\nu) \biggr\},
\end{align}
this implies $\delta_{\xi} \mathcal{S}_{EH}=0$. That is to say, the Einstein-Hilbert action is invariant under global supersymmetry transformations in terms of Weyl spinors.

\bigskip

\end{appendix}

\bibliographystyle{JHEP}
\bibliography{Sugra}

\providecommand{\href}[2]{#2}\begingroup\raggedright\begin{thebibliography}{10}

\bibitem{Feynman:1963ax}
R.~P. Feynman, {\it {Quantum theory of gravitation}},  {\em Acta Phys. Polon.}
  {\bf 24} (1963) 697--722.

\bibitem{DeWitt1967a}
B.~S. DeWitt, {\it Quantum theory of gravity. i. the canonical theory},  {\em
  Phys. Rev.} {\bf 160} (Aug, 1967) 1113--1148.

\bibitem{DeWitt1967b}
B.~S. DeWitt, {\it Quantum theory of gravity. ii. the manifestly covariant
  theory},  {\em Phys. Rev.} {\bf 162} (Oct, 1967) 1195--1239.

\bibitem{DeWitt:1967uc}
B.~S. DeWitt, {\it {Quantum Theory of Gravity. 3. Applications of the Covariant
  Theory}},  {\em Phys. Rev.} {\bf 162} (1967) 1239--1256.

\bibitem{tHooft:1974toh}
G.~'t~Hooft and M.~J.~G. Veltman, {\it {One loop divergencies in the theory of
  gravitation}},  {\em Ann. Inst. H. Poincare Phys. Theor. A} {\bf 20} (1974)
  69--94.

\bibitem{Hooft:2016gir}
G.~'t~Hooft, {\it {Reflections on the renormalization procedure for gauge
  theories}},  {\em Nucl. Phys. B} {\bf 912} (2016) 4--14,
  [\href{http://arxiv.org/abs/1604.06257}{{\tt arXiv:1604.06257}}].

\bibitem{Weinberg:2016kyd}
S.~Weinberg, {\it {Effective field theory, past and future}},  {\em Int. J.
  Mod. Phys. A} {\bf 31} (2016), no.~06 1630007.

\bibitem{Donoghue:1993eb}
J.~F. Donoghue, {\it {Leading quantum correction to the Newtonian potential}},
  {\em Phys. Rev. Lett.} {\bf 72} (1994) 2996--2999,
  [\href{http://arxiv.org/abs/gr-qc/9310024}{{\tt gr-qc/9310024}}].

\bibitem{Donoghue:1994dn}
J.~F. Donoghue, {\it {General relativity as an effective field theory: The
  leading quantum corrections}},  {\em Phys. Rev. D} {\bf 50} (1994)
  3874--3888, [\href{http://arxiv.org/abs/gr-qc/9405057}{{\tt gr-qc/9405057}}].

\bibitem{Bjerrum-Bohr:2002aqa}
N.~E.~J. Bjerrum-Bohr, {\it {Leading quantum gravitational corrections to
  scalar QED}},  {\em Phys. Rev. D} {\bf 66} (2002) 084023,
  [\href{http://arxiv.org/abs/hep-th/0206236}{{\tt hep-th/0206236}}].

\bibitem{Mangano:1990by}
M.~L. Mangano and S.~J. Parke, {\it {Multiparton amplitudes in gauge
  theories}},  {\em Phys. Rept.} {\bf 200} (1991) 301--367,
  [\href{http://arxiv.org/abs/hep-th/0509223}{{\tt hep-th/0509223}}].

\bibitem{Dixon:1996wi}
L.~J. Dixon, {\it {Calculating scattering amplitudes efficiently}},  in {\em
  {Theoretical Advanced Study Institute in Elementary Particle Physics (TASI
  95): QCD and Beyond}}, pp.~539--584, 1, 1996.
\newblock \href{http://arxiv.org/abs/hep-ph/9601359}{{\tt hep-ph/9601359}}.

\bibitem{Bern:1996je}
Z.~Bern, L.~J. Dixon, and D.~A. Kosower, {\it {Progress in one loop QCD
  computations}},  {\em Ann. Rev. Nucl. Part. Sci.} {\bf 46} (1996) 109--148,
  [\href{http://arxiv.org/abs/hep-ph/9602280}{{\tt hep-ph/9602280}}].

\bibitem{Parke:1986gb}
S.~J. Parke and T.~R. Taylor, {\it {An Amplitude for $n$ Gluon Scattering}},
  {\em Phys. Rev. Lett.} {\bf 56} (1986) 2459.

\bibitem{Berends:1987me}
F.~A. Berends and W.~T. Giele, {\it {Recursive Calculations for Processes with
  n Gluons}},  {\em Nucl. Phys. B} {\bf 306} (1988) 759--808.

\bibitem{Elvang:2013cua}
H.~Elvang and Y.-t. Huang, {\it {Scattering Amplitudes}},
  \href{http://arxiv.org/abs/1308.1697}{{\tt arXiv:1308.1697}}.

\bibitem{Bern:1999ji}
Z.~Bern and A.~K. Grant, {\it {Perturbative gravity from QCD amplitudes}},
  {\em Phys. Lett. B} {\bf 457} (1999) 23--32,
  [\href{http://arxiv.org/abs/hep-th/9904026}{{\tt hep-th/9904026}}].

\bibitem{Kawai:1985xq}
H.~Kawai, D.~C. Lewellen, and S.~H.~H. Tye, {\it {A Relation Between Tree
  Amplitudes of Closed and Open Strings}},  {\em Nucl. Phys. B} {\bf 269}
  (1986) 1--23.

\bibitem{Bern:2008qj}
Z.~Bern, J.~J.~M. Carrasco, and H.~Johansson, {\it {New Relations for
  Gauge-Theory Amplitudes}},  {\em Phys. Rev. D} {\bf 78} (2008) 085011,
  [\href{http://arxiv.org/abs/0805.3993}{{\tt arXiv:0805.3993}}].

\bibitem{Bern:2010ue}
Z.~Bern, J.~J.~M. Carrasco, and H.~Johansson, {\it {Perturbative Quantum
  Gravity as a Double Copy of Gauge Theory}},  {\em Phys. Rev. Lett.} {\bf 105}
  (2010) 061602, [\href{http://arxiv.org/abs/1004.0476}{{\tt
  arXiv:1004.0476}}].

\bibitem{Arkani-Hamed:2017jhn}
N.~Arkani-Hamed, T.-C. Huang, and Y.-t. Huang, {\it {Scattering amplitudes for
  all masses and spins}},  {\em JHEP} {\bf 11} (2021) 070,
  [\href{http://arxiv.org/abs/1709.04891}{{\tt arXiv:1709.04891}}].

\bibitem{Diaz-Cruz:2016abv}
J.~L. Diaz-Cruz and B.~O. Larios, {\it {Helicity Amplitudes for massive
  gravitinos in N=1 Supergravity}},  {\em J. Phys. G} {\bf 45} (2018), no.~1
  015002, [\href{http://arxiv.org/abs/1612.04331}{{\tt arXiv:1612.04331}}].

\bibitem{Freedman:1976xh}
D.~Z. Freedman, P.~van Nieuwenhuizen, and S.~Ferrara, {\it {Progress Toward a
  Theory of Supergravity}},  {\em Phys. Rev. D} {\bf 13} (1976) 3214--3218.

\bibitem{Wess:1992cp}
J.~Wess and J.~Bagger, {\em {Supersymmetry and supergravity}}.
\newblock Princeton University Press, Princeton, NJ, USA, 1992.

\bibitem{Freedman:2012zz}
D.~Z. Freedman and A.~Van~Proeyen, {\em {Supergravity}}.
\newblock Cambridge Univ. Press, Cambridge, UK, 5, 2012.

\bibitem{Peskin:1995ev}
M.~E. Peskin and D.~V. Schroeder, {\em {An Introduction to quantum field
  theory}}.
\newblock Addison-Wesley, Reading, USA, 1995.

\bibitem{Arkani-Hamed:2012zlh}
N.~Arkani-Hamed, J.~L. Bourjaily, F.~Cachazo, A.~B. Goncharov, A.~Postnikov,
  and J.~Trnka, {\em {Grassmannian Geometry of Scattering Amplitudes}}.
\newblock Cambridge University Press, 4, 2016.

\bibitem{Britto:2004ap}
R.~Britto, F.~Cachazo, and B.~Feng, {\it {New recursion relations for tree
  amplitudes of gluons}},  {\em Nucl. Phys. B} {\bf 715} (2005) 499--522,
  [\href{http://arxiv.org/abs/hep-th/0412308}{{\tt hep-th/0412308}}].

\bibitem{Britto:2005fq}
R.~Britto, F.~Cachazo, B.~Feng, and E.~Witten, {\it {Direct proof of tree-level
  recursion relation in Yang-Mills theory}},  {\em Phys. Rev. Lett.} {\bf 94}
  (2005) 181602, [\href{http://arxiv.org/abs/hep-th/0501052}{{\tt
  hep-th/0501052}}].

\bibitem{Benincasa:2007xk}
P.~Benincasa and F.~Cachazo, {\it {Consistency Conditions on the S-Matrix of
  Massless Particles}},  \href{http://arxiv.org/abs/0705.4305}{{\tt
  arXiv:0705.4305}}.

\bibitem{Cheung:2017pzi}
C.~Cheung, {\em {TASI Lectures on Scattering Amplitudes}}, pp.~571--623.
\newblock World Scientific, 2018.
\newblock \href{http://arxiv.org/abs/1708.03872}{{\tt arXiv:1708.03872}}.

\bibitem{Schwartz:2014sze}
M.~D. Schwartz, {\em {Quantum Field Theory and the Standard Model}}.
\newblock Cambridge University Press, 3, 2014.

\bibitem{Bjerrum-Bohr:2010eeh}
N.~E.~J. Bjerrum-Bohr and O.~T. Engelund, {\it {Gravitino Interactions from
  Yang-Mills Theory}},  {\em Phys. Rev. D} {\bf 81} (2010) 105009,
  [\href{http://arxiv.org/abs/1002.2279}{{\tt arXiv:1002.2279}}].

\bibitem{Holstein:2006bh}
B.~R. Holstein, {\it {Graviton Physics}},  {\em Am. J. Phys.} {\bf 74} (2006)
  1002--1011, [\href{http://arxiv.org/abs/gr-qc/0607045}{{\tt gr-qc/0607045}}].

\bibitem{Moroi:1995fs}
T.~Moroi, {\it {Effects of the gravitino on the inflationary universe}},  other
  thesis, Tohoku University, 3, 1995.

\bibitem{Diaz-Cruz:2019xjb}
J.~L. Diaz-Cruz and B.~O. Larios, {\it {Very long-lived Stop NLSP in MSSM
  scenarios with Gravitino LSP}},  \href{http://arxiv.org/abs/1901.06352}{{\tt
  arXiv:1901.06352}}.

\bibitem{Das:1976ct}
A.~K. Das and D.~Z. Freedman, {\it {Gauge Quantization for Spin 3/2 Fields}},
  {\em Nucl. Phys. B} {\bf 114} (1976) 271--296.

\bibitem{Kleiss:1985yh}
R.~Kleiss and W.~J. Stirling, {\it {Spinor Techniques for Calculating p anti-p
  ---\ensuremath{>} W+- / Z0 + Jets}},  {\em Nucl. Phys. B} {\bf 262} (1985)
  235--262.

\bibitem{Hagiwara:1985yu}
K.~Hagiwara and D.~Zeppenfeld, {\it {Helicity Amplitudes for Heavy Lepton
  Production in e+ e- Annihilation}},  {\em Nucl. Phys. B} {\bf 274} (1986)
  1--32.

\bibitem{Badger:2023eqz}
S.~Badger, J.~Henn, J.~C. Plefka, and S.~Zoia, {\it {Scattering Amplitudes in
  Quantum Field Theory}},  {\em Lect. Notes Phys.} {\bf 1021} (2024) pp.,
  [\href{http://arxiv.org/abs/2306.05976}{{\tt arXiv:2306.05976}}].

\bibitem{Diaz-Cruz:2015oie}
J.~L. Diaz-Cruz, B.~L. Lopez, O.~Meza-Aldama, and J.~R. Perez, {\it {Weyl
  spinors and the helicity formalism}},  {\em Rev. Mex. Fis. E} {\bf 61} (2015)
  104, [\href{http://arxiv.org/abs/1511.07477}{{\tt arXiv:1511.07477}}].

\bibitem{Xu:1986xb}
Z.~Xu, D.-H. Zhang, and L.~Chang, {\it {Helicity Amplitudes for Multiple
  Bremsstrahlung in Massless Nonabelian Gauge Theories}},  {\em Nucl. Phys. B}
  {\bf 291} (1987) 392--428.

\bibitem{Carrasco:2015iwa}
J.~J.~M. Carrasco, {\it {Gauge and Gravity Amplitude Relations}},  in {\em
  {Theoretical Advanced Study Institute in Elementary Particle Physics}:
  {Journeys Through the Precision Frontier: Amplitudes for Colliders}},
  pp.~477--557, WSP, 2015.
\newblock \href{http://arxiv.org/abs/1506.00974}{{\tt arXiv:1506.00974}}.

\bibitem{ReyesPerez:2021amr}
J.~Reyes~P\'erez, {\it {M\'etodos de amplitudes en QFT supersim\'etricas}},
  Master's thesis, FCFM-BUAP, 12, 2021.

\bibitem{Fierz:1939ix}
M.~Fierz and W.~Pauli, {\it {On relativistic wave equations for particles of
  arbitrary spin in an electromagnetic field}},  {\em Proc. Roy. Soc. Lond. A}
  {\bf 173} (1939) 211--232.

\bibitem{Rarita:1941mf}
W.~Rarita and J.~Schwinger, {\it {On a theory of particles with half integral
  spin}},  {\em Phys. Rev.} {\bf 60} (1941) 61.

\bibitem{Dreiner:2008tw}
H.~K. Dreiner, H.~E. Haber, and S.~P. Martin, {\it {Two-component spinor
  techniques and Feynman rules for quantum field theory and supersymmetry}},
  {\em Phys. Rept.} {\bf 494} (2010) 1--196,
  [\href{http://arxiv.org/abs/0812.1594}{{\tt arXiv:0812.1594}}].

\bibitem{Bailin:1994qt}
D.~Bailin and A.~Love, {\em {Supersymmetric Gauge Field Theory and String
  Theory}}.
\newblock Taylor \& Francis, 1994.

\bibitem{Kirchbach:2002nu}
M.~Kirchbach and D.~V. Ahluwalia, {\it {Space-time structure of massive
  gravitino}},  {\em Phys. Lett. B} {\bf 529} (2002) 124--131,
  [\href{http://arxiv.org/abs/hep-th/0202164}{{\tt hep-th/0202164}}].

\bibitem{Napsuciale:2006wr}
M.~Napsuciale, M.~Kirchbach, and S.~Rodriguez, {\it {Spin 3/2 Beyond the
  Rarita-Schwinger Framework}},  {\em Eur. Phys. J. A} {\bf 29} (2006)
  289--306, [\href{http://arxiv.org/abs/hep-ph/0606308}{{\tt hep-ph/0606308}}].

\end{thebibliography}\endgroup


\end{document}